\documentclass[floatfix,prd,aps,twocolumn,letterpaper,lengthcheck,superscriptaddress,showpacs,amssymb,amsmath,amsfonts,aps,altaffilletter,nofootinbib,nopreprintnumbers,showpacs,longbibliography,superscriptaddress]{revtex4-1}

\usepackage{amsfonts}
\usepackage{amsmath}
\usepackage{amssymb}
\usepackage{bm}
\usepackage{dcolumn}
\usepackage{graphicx}
\usepackage{graphics}
\usepackage[latin1]{inputenc}
\usepackage{latexsym}
\usepackage{rotating}
\usepackage{hyperref}
\usepackage{xspace} 
\usepackage[usenames]{color}
\usepackage{mathrsfs}
\usepackage{multirow}

\usepackage{subfigure}

\usepackage{longtable}
\usepackage{pict2e}
\usepackage[usenames,dvipsnames]{xcolor}
\usepackage{url}

\allowdisplaybreaks

\definecolor {darkgreen}{rgb}{0.2,0.7,0.2}
\definecolor{purple}{rgb}{0.5,0,0.5}

\newcommand{\bea}{\begin{eqnarray}}
\newcommand{\eea}{\end{eqnarray}}
\newcommand{\beq}{\begin{equation}}
\newcommand{\eeq}{\end{equation}}

\begin{document}

\title{Eccentric Binary Black Holes with Spin \\ via the Direct Integration of the Post-Newtonian Equations of Motion}

\author{Brennan~Ireland}
\email{bmi5921@rit.edu}
\affiliation{Center for Computational Relativity and Gravitation, 
School of Mathematical Sciences, and School of Physics and Astronomy, Rochester Institute of
Technology, Rochester, New York 14623, USA.}

\author{Ofek~Birnholtz}
\email{ofek@mail.rit.edu}
\affiliation{Center for Computational Relativity and Gravitation, 
School of Mathematical Sciences, and School of Physics and Astronomy, Rochester Institute of
Technology, Rochester, New York 14623, USA.}

\author{Hiroyuki~Nakano}
\email{hinakano@law.ryukoku.ac.jp}
\affiliation{Center for Computational Relativity and Gravitation, 
School of Mathematical Sciences, and School of Physics and Astronomy, Rochester Institute of
Technology, Rochester, New York 14623, USA.}
\affiliation{Faculty of Law, Ryukoku University, Kyoto 612-8577, Japan.}

\author{Eric~West}
\affiliation{University of Minnesota Duluth, Duluth, Minnesota 55812, USA.}

\author{Manuela~Campanelli}
\affiliation{Center for Computational Relativity and Gravitation, 
School of Mathematical Sciences, and School of Physics and Astronomy, Rochester Institute of
Technology, Rochester, New York 14623, USA.}

\begin{abstract}

We integrate the third and a half post-Newtonian equations of motion
for a fully generic binary black hole system,
	allowing both for non-circular orbits,
	and for one or both of the black holes to spin, in any orientation.
Using the second post-Newtonian order expression
beyond the leading order quadrupole formula,
we study the gravitational waveforms
	produced from such systems. 
Our results are validated by comparing to Taylor T4
	in the aligned-spin circular cases,
	and the additional effects and modulations introduced
	by the eccentricity and the spins are analyzed.
We use the framework to evaluate the evolution of eccentricity,
	and trace its contributions to source terms corresponding to the different definitions.
Finally, we discuss how this direct integration equations-of-motion code
may be relevant to existing and upcoming gravitational wave detectors,
showing fully generic, precessing, eccentric gravitational
waveforms from a fiducial binary system
with the orbital plane and spin precession, and the eccentricity reduction.

\end{abstract}

\date{2019-04-05}

\maketitle


\section{Introduction}

The direct detections of gravitational waves (GWs)
\cite{GWdetect,TheLIGOScientific:2016pea,LIGOScientific:2018mvr}
by the LIGO/VIRGO Collaboration~\cite{AdvLIGO,AdvVIRGO,
Abbott:2018LRR....21....3A}
have launched the era of multi-messenger astrophysics,
both in providing a new window on events such as binary neutron star (BNS)
mergers~\cite{GW170817:NSmerger},
and by opening for study a completely new field of
astrophysical events previously invisible to us,
binary black hole (BBH) mergers \cite{LIGOScientific:2018mvr},
meriting the Nobel Prize in Physics in 2017.

GWs from eccentric binaries \cite{Gopakumar:2002PhRvD..65h4011G,
Yunes:2009PhRvD..80h4001Y,Huerta:2014PhRvD..90h4016H,
Hannam:2014PhRvL.113o1101H,Taracchini:2014PhRvD..89f1502T,
Tanay:2016PhRvD..93f4031T, Huerta:2016rwp, Huerta:2017kez,
Hinder:2017sxy,Lower:2018seu,Huerta:2019oxn} have become
	an important topic of study as 
	the next observing runs of the Advanced Laser Interferometer
	Gravitational-wave Observatory (aLIGO) and VIRGO approach.
While it is not expected that LIGO/VIRGO sources in isolated binaries will
have significant eccentricity in band,
binaries in dense stellar clusters \cite{Merritt_dyn} may, however, 
retain significant eccentricity from dynamical interactions
prior to entry into the LIGO/VIRGO band.
Recent simulations of globular clusters indicate that a distinct population
of compact binaries exist
(from Refs.~\cite{Rodriguez:2018PhRvL.120o1101R,Rodriguez:2016PhRvD}, about 3\% of binaries) and
enter the LIGO/VIRGO band (canonically taken to be $10$\,Hz as the lower bound)
with significant eccentricity
($e > 0.1$)~\cite{Rodriguez:2018PhRvL.120o1101R, Antonini:2016ApJ...831..187A}.
These results add further motivation, 
for if binaries can form and merge in this way,
then there will be a minority (but a distinct minority),
that will enter with eccentricity, 
and which may be missed without templates for the match filtering.

In addition, a large fraction of binaries in globular clusters will have significant 
eccentricity ($\sim 50\%$ of in-cluster mergers) and will be detectable
for the entirety of the LISA band
($\sim 10^{-3}$ -- $10^{-1}$\,Hz)~\cite{Breivik:2016ApJ...830L..18B,
DOrazio:2018arXiv180506194D}. 
Both 2-body mergers (highly eccentric in cluster mergers 
in between single binary interactions) and 
3-body mergers (when a BBH forms with such high eccentricity
that it is essentially a GW capture before it can interact with a third body)
will occur in clusters.
LISA will be able to measure the 2-body mergers, but 
not the 3-body mergers~\cite{Samsing:2018arXiv180406519S}. 
Current work is being carried out to see if detection 
of eccentric sources will be possible with the proposed space-based detector,
and current prospects look promising~\cite{DOrazio:2018arXiv180506194D,
Breivik:2016ApJ...830L..18B, Sesana:2016PhRvL.116w1102S}.

In this paper,
we develop and calculate the GW waveforms and orbital dynamics from 
generically spinning, eccentric BBHs. 
We use the Lagrangian formulation of the post-Newtonian (PN) equations
of motion (EOM) in the harmonic gauge
for the generation of precessing,
eccentric GW signatures~\cite{Blanchet:1995fg, Blanchet:1995fr,
Blanchet:1996wx, Blanchet:1998vx, Blanchet:2002av, Will:2005sn,
Faye:2006gx, Blanchet:2006gy,
Will:2011PNAS..108.5938W, Marsat:2012fn, Bohe:2012mr, 
Blanchet:2013haa, Marsat:2014xea, Bohe:2015ana, Bernard:2018PhRvD..97d4037B}.
Our approach allows us to use any spin values, mass ratios, and eccentricities, 
without restricting to planar orbits or co-precessing frames,
so long as the binary has a large enough separation
(around a periapsis passage of $r\sim10M$), 
such that the underlying PN theory does not break down~\cite{Poisson:1995vs,Mino:1997bx,Yunes:2008tw,Zhang:2011vha,Sago:2016xsp,Fujita:2017wjq}.
This approach offers a major step forward
as a way to generate eccentric,
precessing GW waveforms in a direct and straightforward way,
without carrying any of the additional restrictive assumptions 
of quasi-ellipticity or adiabaticity that many current waveform generators use.
Furthermore, we can give more precise waveforms
since we do not ignore any timescale effect
(see, e.g., Ref.~\cite{Klein:2018ybm}).

This work is an extension of work first done by Lincoln and Will~\cite{WillandLincoln:1990PhRvD..42.1123L},
and of integrations as in Refs.~\cite{Pati:2000vt,Pati:2002ux}.
In more recent works, Refs.~\cite{Levin:2011CQG}
and~\cite{Csizmadia:2012CQGra..29x5002C}
used this framework to 3.5PN to generate eccentric binaries
in the PN harmonic gauge, and calculate the relevant waveforms. 
The research focus of this project is to calculate the orbital quantities
and GW waveforms from generic binaries with the requisite accuracy
to enable potential future observations from GW detectors.
This paper details the methodology, and extends the previous work
done by other authors
by giving quantitative comparisons to other known PN methods for the first time.

This Lagrangian formulation has larger applications as well. 
Developing a general EOM that can handle arbitrarily precessing and eccentric BBHs
means that we can apply this EOM to the precessing spacetime
developed in our previous work~\cite{Nakano:2016CQGra..33x7001N}.
We must have an accurate EOM for the general 
precessing BBH that can handle the orbital plane precession of the binary
due to spin coupling with the orbit~\cite{Owen:1997ku},
and also the individual spins precessing in order to evolve the BBH. 
The Lagrangian formulation excels at all of these,
and can directly be used for this evolution. 

The regime of final inspiral to plunge and merger can only be modeled
by numerical relativity~\cite{Pretorius:2005gq, Campanelli:2005dd,
Baker:2005vv, Campanelli:2006uy, Campanelli:2006PhRvD,
Campanelli:2007ew, Campanelli:2007apj, Lousto:2014ida, Lousto:2015uwa}.
This does, however, provide us a unique
opportunity to use the numerical relativity regime; we can stitch our PN 
evolution onto the beginning of NR simulations,
and thus create a full waveform model for the binary using
hybridization of waveforms~\cite{Campanelli:2009PhRvD..79h4010C, Ajith:2012tt}. 

This paper is organized as follows. 
Sec.~\ref{sec:PN_form} gives an overview of the PN 
EOMs for generic spins and eccentricities.
Sec.~\ref{sec:ecc_spin_binaries} shows the power 
and flexibility of the EOM code, and demonstrates our main results 
with orbital and waveform quantities for some fiducial systems
with eccentricity and spins.
Sec.~\ref{sec:comp_T4_EOM} quantifies the comparisons of the EOM code
with the Taylor T4 method, the initial conditions for quasi-circular orbits,
including orbital frequency comparison, and waveform overlaps.
In Sec.~\ref{sec:generictest}, 
we discuss a simple test of orbital eccentricity by using the EOM.
Sec.~\ref{sec:discussion} contains discussion, conclusions,
and future work. 

Throughout this paper, we use the geometric unit system, where $c=G=1$,
with the useful conversion factor 
$1 M_{\odot} = 1.477 \; {\rm{km}} = 4.926 \times 10^{-6} \; {\rm{s}}$,
although we will keep some $1/c$ factors to count PN orders.

\section{post-Newtonian Formulation} \label{sec:PN_form}

\subsection{Orbital Position and Spin Equations of Motion}

The general case of unaligned spins (i.e., spins that are neither 
aligned nor anti-aligned with the orbital angular momentum) leads to 
precession of the individual spins as well as precession of the orbital
angular momentum vector, and therefore precession of the orbital plane.
To describe this complicated motion requires two sets of evolution 
equations: one set describing the positions (trajectories) of the point 
masses, and one set describing the evolution of the spin of each point 
mass. We will review the position equations first, then the precession 
equations. There are several PN papers that go through
this in detail. See Ref.~\cite{Blanchet:2013haa} for the non-spinning terms, 
and Refs.~\cite{Marsat:2012fn, Bohe:2012mr} for the spin dependent terms.

The system comprises of two bodies ($a=1,2$),
each described by a mass $m_a$, a spin $\vec{S}_a$,
a spin precession vector $\vec{\Omega}_a$,
and position and velocity vectors $\vec{r}_a$, $\vec{v}_a$.
We also define the total combinations
$M=m_1+m_2$ for the total mass,
$\eta=m_1m_2/M^2$ for the symmetric mass ratio,
$\mu=M\eta$ for the reduced mass, and
\bea
\vec{S} = \vec{S}_1 + \vec{S}_2 \,,
\eea
for the total spin, as well as the relative combinations
$\vec{r}=\vec{r}_1-\vec{r}_2 = r\hat{n}$ (where $r=|\vec{r}|$)
denoting the relative separation vector,
$\vec{v}=\vec{v}_1-\vec{v}_2$ for the relative velocity,
and the relative spin difference $\vec{\Sigma}$, given by
\bea
\frac{\vec{\Sigma}}{M} = \frac{\vec{S}_2}{m_2} - \frac{\vec{S}_1}{m_1} \,.
\eea

The schematic EOM for the orbital positions,
in terms of relative variables and in the center-of-mass frame,
then takes the form:
\begin{eqnarray} \label{direct_EOM}
\frac{\mathrm{d}\vec{v}}{\mathrm{d}t} &=& 
-\frac{M}{r^2} \left[ (1 + {\mathcal{A}})\hat{n} 
+ {\mathcal{B}}\vec{v} \right] \nonumber \\
&&
+ {\mathcal{C}}_{1}\,(\hat{n}\times\vec{S}) 
+ {\mathcal{C}}_{2}\,(\hat{n}\times\vec{\Sigma})
\cr &&
+ {\mathcal{C}}_{3}\,(\vec{S}\times\vec{v}) 
+ {\mathcal{C}}_{4}\,(\vec{\Sigma}\times\vec{v}) \,,
\end{eqnarray}
with the non-spinning components of these $\mathcal{A}$ and $\mathcal{B}$
terms defined in Ref.~\cite{Blanchet:2013haa},
and the spin terms and expressions for ${\mathcal{C}}_{1}$,
${\mathcal{C}}_{2}$, ${\mathcal{C}}_{3}$, and ${\mathcal{C}}_{4}$
defined in Refs.~\cite{Faye:2006gx, Blanchet:2006gy, Marsat:2012fn, Bohe:2012mr}.

The corresponding precession equations from Ref.~\cite{Bohe:2012mr}
take the form:
\begin{subequations}
\begin{eqnarray}
\frac{\mathrm{d}\vec{S}}{\mathrm{d}t} &=& 
\frac{m_1\vec{\Omega}_1 + m_2\vec{\Omega}_2}{M}\times \vec{S}
+ \eta(\vec{\Omega}_2 - \vec{\Omega}_1)\times \vec{\Sigma} \,,
\label{sEOM.1}
\\
\frac{\mathrm{d}\vec{\Sigma}}{\mathrm{d}t} &=& 
\frac{m_2\vec{\Omega}_1 + m_1\vec{\Omega}_2}{M}\times \vec{\Sigma}
+ (\vec{\Omega}_2 - \vec{\Omega}_1)\times \vec{S} \,.
\label{sEOM.2}
\end{eqnarray}
\end{subequations}
Following Refs.~\cite{Will:2005sn, Blanchet:2006gy, Blanchet:2013haa},
we use the Tulczyjew spin supplementary condition (SSC)
to define a spin vector with conserved Euclidean norm.
For this SSC, the higher order spin-orbit terms have been derived 
in Refs.~\cite{Faye:2006gx, Blanchet:2006gy, Marsat:2012fn,
Bohe:2012mr,Blanchet:2013haa}. 
The next-to-leading order spin-spin terms were derived 
in Ref.~\cite{Bohe:2015ana}, and the leading order cubic-in-spin terms were derived 
in Ref.~\cite{Marsat:2014xea}.
To date, leading order quartic- and quintic-in-spin contributions 
to the EOM have not been derived in PN harmonic coordinates with this SSC.
See Appendix~\ref{sec:sEOM} for some details. 

Up to 3.5PN~\footnote{A PN order $N$ is said to be a term of order $(v/c)^{2N}$.}
order in this formalism, and for maximal spins
($|\vec{\chi}_\mathrm{a}|\simeq1$ where
$\vec{\chi}_\mathrm{a}=\vec{S}_\mathrm{a}/m_\mathrm{a}^2$), 
spin-orbit effects contribute to the EOM at 1.5PN, 2.5PN, and 3.5PN. 
Spin-spin effects contribute at 2PN and 3PN. 
Cubic-in-spin effects start from 3.5PN. 
Quartic- and higher order in-spin effects are beyond 3.5PN. 

Equations~\eqref{sEOM.1} and \eqref{sEOM.2}, along with the acceleration 
equation~\eqref{direct_EOM} comprise the EOMs of the two-body 
system. For a given set of initial conditions, this system of ODE's is 
easily solved numerically~\footnote{We have used \textit{Mathematica}'s
built-in RK4 solver to solve the EOMs.}.

We close this section with some brief comments about radiation reaction.
Through 3.5PN order (inclusive), radiation reaction effects arise 
{\it exclusively} in the non-spin part of $\mathcal{B}$. 
Radiation reaction effects first 
appear in the non-spin part of $\mathcal{A}$ at 6PN. Spin effects enter 
radiation reaction at 4PN, and radiation reaction effects 
enter spin contributions at 4PN. Furthermore, for quasi-circular orbits,
the non-spin part of $\mathcal{B}$ contains {\it only} radiation reaction
effects. In that case, radiation reaction is completely controlled to 3.5PN
by turning on/off the non-spin part of $\mathcal{B}$.

\subsection{Generation of Gravitational Radiation}

To calculate the polarization of GWs, $h_+$ and $h_{\times}$, we need to define 
the principle axes for the GWs. In particular, we define 
$\hat{N}$ which is the radial direction to the observer, $\hat{p}$, which lies along 
the line of nodes (for our purposes we can set to be along the $y$ axis), and 
$\hat{q}$ which is orthogonal to $\hat{N}$ and $\hat{p}$~\cite{Will:1996zj}.
Note that $\hat{N} \neq \hat{p}\times\hat{q}$.

\begin{figure}[t!]
\includegraphics[width= \columnwidth,clip=true]{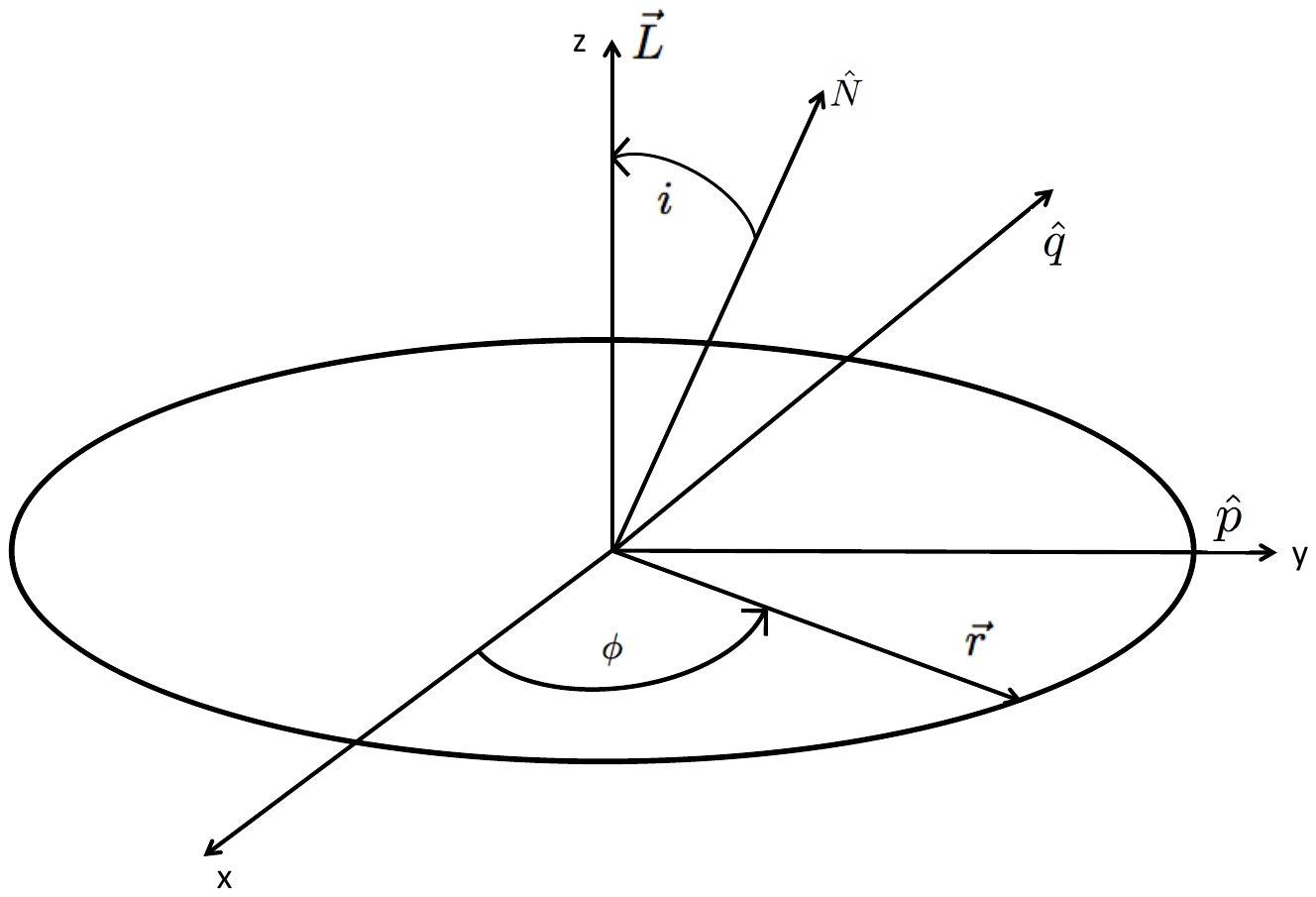}
\caption{The orbit-adapted vectors for GW polarizations. $\hat{N}$ is the 
direction to the observer, $\hat{p}$ is along the line of nodes,
and $\vec{q} = \hat{N} \times \hat{p}$.
The phase of the binary $\phi$ is defined in a right-handed sense.
For the initial conditions,
the binary starts on the $x$ axis, and the phase is defined from there.
The Newtonian angular momentum $\vec{L} = \mu \vec{r} \times \vec{v}$ defines the inclination $i$.}
\label{orbit_adapted_vecs}
\end{figure}

This allows us to define the 
inclination $i$, and the phase $\phi$, with respect
to the Newtonian angular momentum 
$\vec{L} = \mu \vec{r} \times \vec{v}$
(see Fig.~\ref{orbit_adapted_vecs}).

For the GW calculation, we use the 2PN order formula
beyond the leading order quadrupole approximation by following Ref.~\cite{Will:1996zj}:
\bea
h^{ij} &=& \frac{2 \mu}{R} \bigg[ Q^{ij} + P^{1/2} Q^{ij} + P Q^{ij} + P Q^{ij}_{\rm SO} 
\cr && \qquad 
+ P^{3/2} Q^{ij}
+ P^{3/2} Q^{ij}_{\rm SO} + P^2 Q^{ij}
\cr && \qquad 
+ P^2 Q^{ij}_{\rm SO} + P^2 Q^{ij}_{\rm SS} \bigg]_{\rm TT} \,, 
\label{eq:2PNQPwave}
\eea
where $R$ denotes
the distance between the observer and the binary,
and the individual PN terms are broken up into
non-spinning, spin-orbit (SO), and spin-spin (SS) 
contributions. 
For example, $P^{3/2}Q_{\rm SO}^{ij}$ is the 1.5PN SO contribution,
and TT means the transverse-traceless part.
We have implemented all of the non-spinning contributions 
to the quadrupole moment up to 2PN, in order to be in agreement with 
the GW calculations for Taylor T4. 

For the Taylor T4 GW waveforms, we follow Ref.~\cite{Blanchet:2013haa}, which 
expands the waveforms $h_+$ and $h_{\times}$ into PN orders as powers of the 
frequency variable $x = (M \Omega)^{2/3}$,
\bea
h_{+,\times} &=& \frac{2 \mu x}{R} \sum^{+ \infty}_{p=0} x^{p/2} H_{(p/2) +,\times}(\psi, \cos i, \sin i; \ln x) 
\cr && + O \bigg(\frac{1}{R^2}\bigg) \,,
\eea
to the desired PN order. The phase variable $\psi$ is related to the binary orbital phase 
$\phi$ by the auxiliary phase variable~\cite{Blanchet:2013haa}:
\bea
\psi = \phi - 2 M_{\rm ADM} \Omega \ln \,\bigg( \frac{\Omega}{\Omega_0} \bigg) \,.
\eea 
The constant frequency $\Omega_0$ can be chosen at will, and for our analysis
is set to the one related to $10$\,Hz
(chosen naively as the entry frequency into the LIGO/VIRGO band). The mass variable in the 
GW calculation is the ADM mass of the binary:
\bea
M_{\rm ADM} = M \bigg[ 1 - \frac{\eta}{2} \gamma + \frac{\eta}{8} (1 - \eta) \gamma^2 + O\bigg( \frac{1}{c^6}\bigg) \bigg] \,,
\eea
where
$\gamma = M/r$ is an expression of the PN parameter of the system. 
The $H_{(p/2) +,\times}(\psi, \cos i, \sin i; \ln x)$ terms are
the specific expansion coefficients, 
and are dependent on the auxiliary phase 
variable $\psi$, and the inclination 
of the binary $i$. The log terms of the frequency variable first occur at 3PN
in the GW waveform.
Explicit expressions for $H_{(p/2) +,\times}$
appear in Eqs.~(322)
and (323) of Ref.~\cite{Blanchet:2013haa},
which can be easily computed from the output of our EOM integrator.

\section{Eccentric, Precessing Binaries}\label{sec:ecc_spin_binaries}

We now demonstrate the power of our framework for the most generic systems of BBHs,
ones with non-unity mass ratios, spin orientations, and eccentricities. 
Since this parameter space is very large, we use a fiducial example for a 
demonstration of the orbital and waveform quantities,
and know that the formalism is generic for any set of mass, spin,
and eccentricity parameters.

For the following example, we follow Ref.~\cite{Rodriguez:2018PhRvL.120o1101R}
to draw a fiducial binary from the histograms that globular cluster 
simulations produce. Other than picking physically 
relevant parameters, we do not restrict ourselves to any particular parameters.

The system that we chose to evaluate is a binary with an initial periapsis 
passage of $35M$, an initial value eccentricity picked to be $0.4$,
a mass ratio $q = 3/2$, and 
dimensionless spin parameter values of
$\chi_1 = (-0.3, 0, 0)$ and $\chi_2 = (0, 0, 0.3)$.
The initial periapsis passage and initial eccentricity were selected such that
a binary with masses comparable to detected LIGO sources (we use a total 
mass of 25 solar masses in the analysis of the GW waveform later 
to dimensionalize the units) would fall just into the LIGO band at periapsis.
The chosen spin parameters have low spin to lie in the physically relevant 
globular cluster results, and are initially strongly misaligned
to give lots of spin-spin and spin-orbit precession.

The extrinsic parameters for waveform production that we chose
for this fiducial system are optimized for ease.
The distance from the source we set to be $500$\,Mpc 
(a redshift of $\sim 0.1$, small enough not to need to take cosmological effects
into account), with the orientation set to $(i, \varpi, \Omega) = (0, 0, 0)$
where $i$, $\varpi$ and $\Omega$ are the extrinsic orbital parameters,
and denote the inclination angle, argument of periapsis
and longitude of the ascending node, respectively.
This is referred to as the optimal orientation, where the binary
is face on (inclination equal to zero), the argument of periapsis is set to zero,
and the longitude of the ascending node is equal to zero (the binary is not tilted 
with respect to the observer). These 13 parameters are enough to completely specify
the binary that we are describing (though it will not give sky localization,
as this set of parameters is only for one detector,
and a second detector would be needed for localization purposes).
For a real source, it is a straightforward 
generalization to put in the detector antenna patterns. 

The final thing that we will need for the analysis 
is the eccentricity definition which we use for generic eccentricities.
This is given as:
\bea \label{geom_ecc_def}
e = \frac{r_{\rm max} - r_{\rm min}}{r_{\rm max} + r_{\rm min}} \,,
\eea
which is a workable definition of eccentricity
for the orbits~\cite{Levin:2011CQG, Csizmadia:2012CQGra..29x5002C}
in the situation of non-negligible eccentricity.
It is noted that this definition relies on the position of the binary
at different points in the orbit; 
therefore, it is not an instantaneous definition at a point,
but is instead averaged over the orbit. 

\begin{figure}[tb!]
\includegraphics[width= \columnwidth,clip=true]{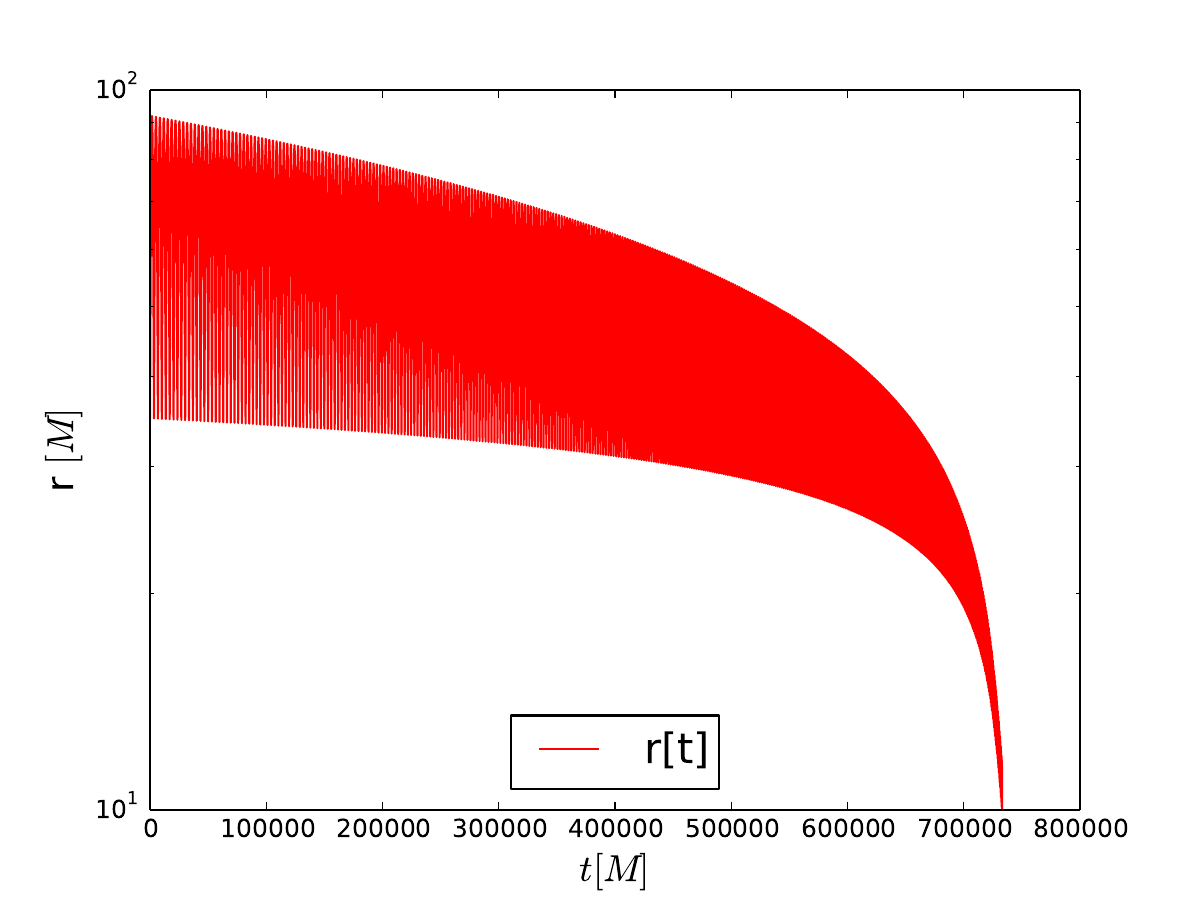}
\caption{
The orbital separation as a function of time for the fiducial system
that we lay out below. 
The binary has an initial periapsis passage of $35M$,
an initial eccentricity of $0.4$, a mass ratio $q = 3/2$, and 
initial dimensionless spin parameter values of
$\chi_1 = (-0.3, 0, 0)$ and $\chi_2 = (0, 0, 0.3)$.
The orbital eccentricity as the binary evolves is
calculated from Eq.~\eqref{geom_ecc_def}.
}
\label{fig:sepvst}
\end{figure}

\begin{figure}[tb!]
\includegraphics[width= \columnwidth,clip=true]{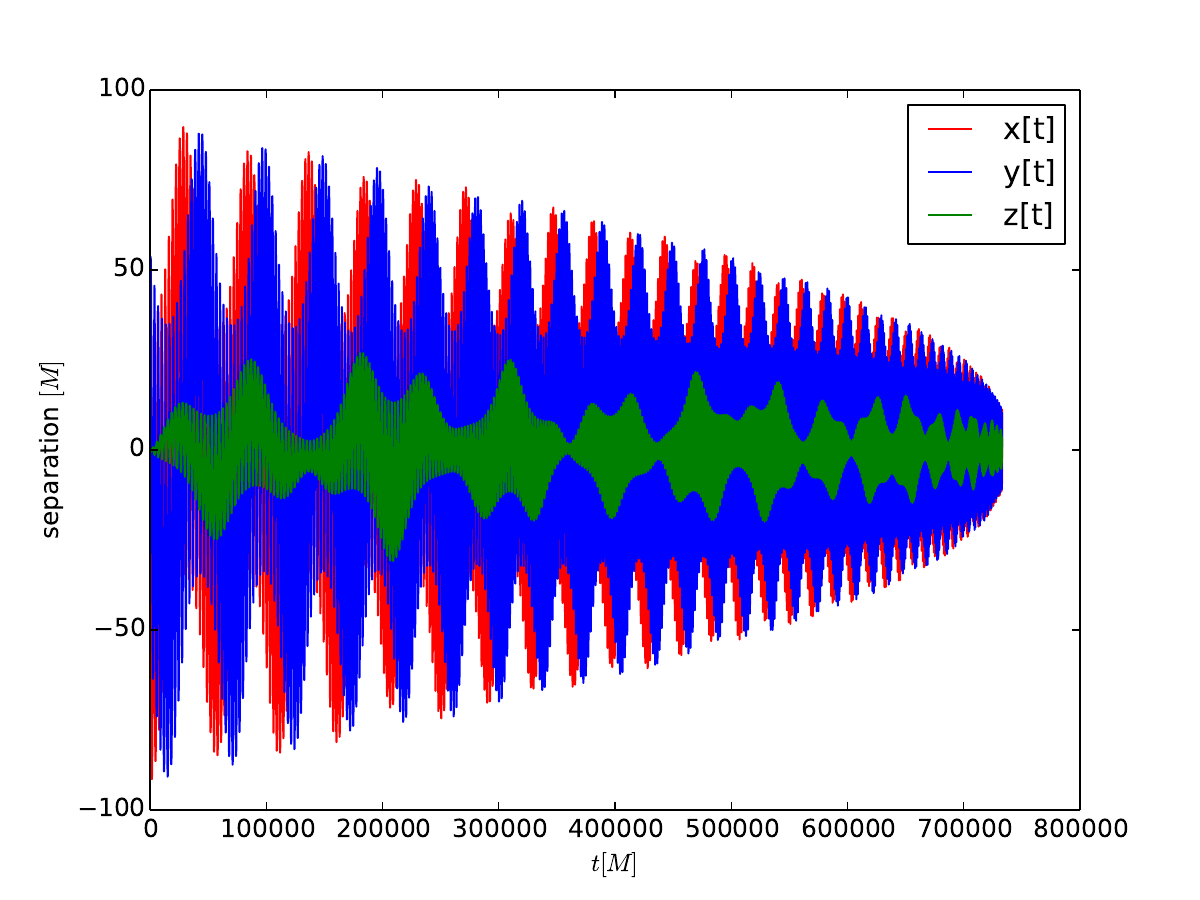}
\caption{The individual components of the trajectory as a function of time 
in the center of mass variables for the fiducial binary system
that we outlined below. 
The binary has an initial periapsis 
passage of $35M$, an initial eccentricity of $0.4$, a mass ratio $q = 3/2$, and 
initial dimensionless spin parameter values of
$\chi_1 = (-0.3, 0, 0)$ and $\chi_2 = (0, 0, 0.3)$.
Note here the spin-orbit coupling giving rise 
to a non zero $z$ component of the orbital motion.}
\label{fig:Ecc_prec_orbits}
\end{figure}

\begin{figure*}[htb!]
\includegraphics[width= 0.67 \columnwidth,clip=true]{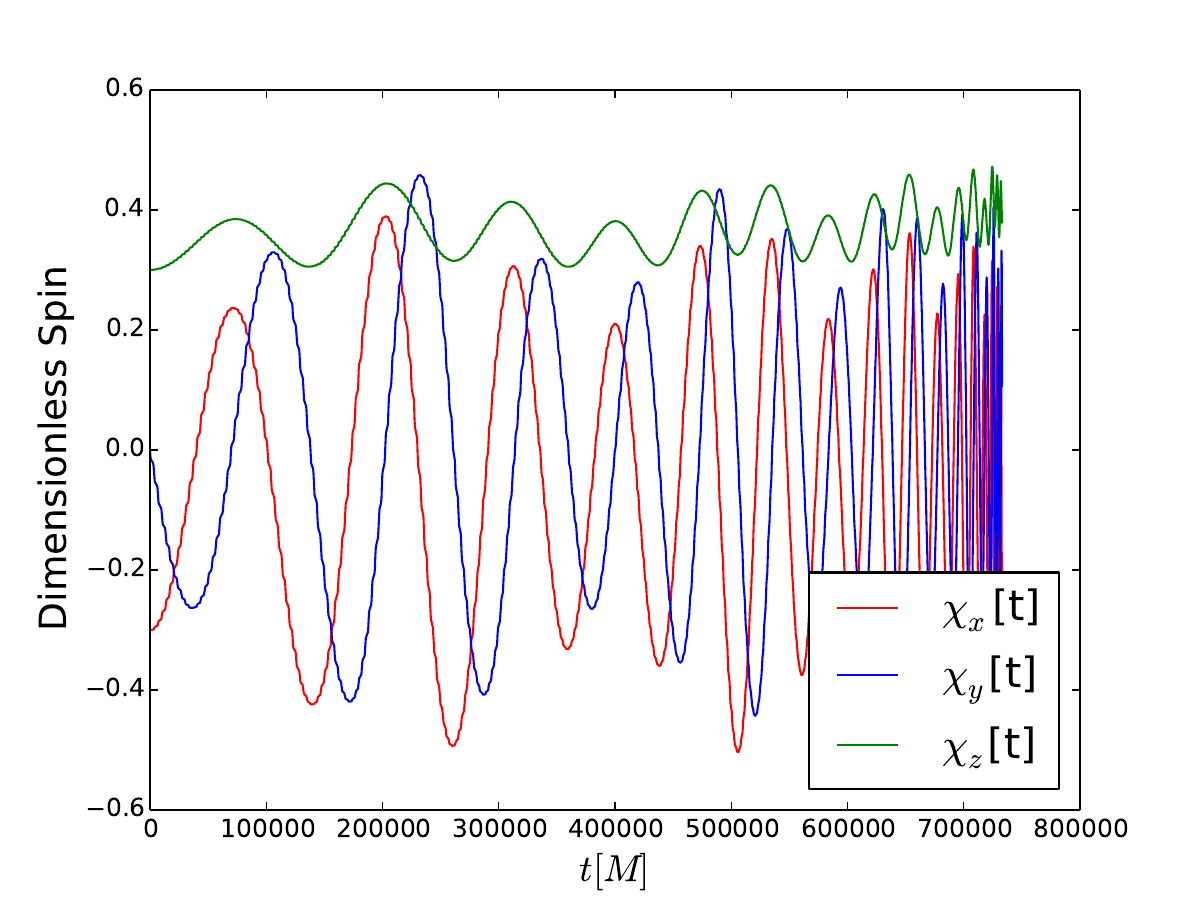}
\includegraphics[width= 0.67 \columnwidth,clip=true]{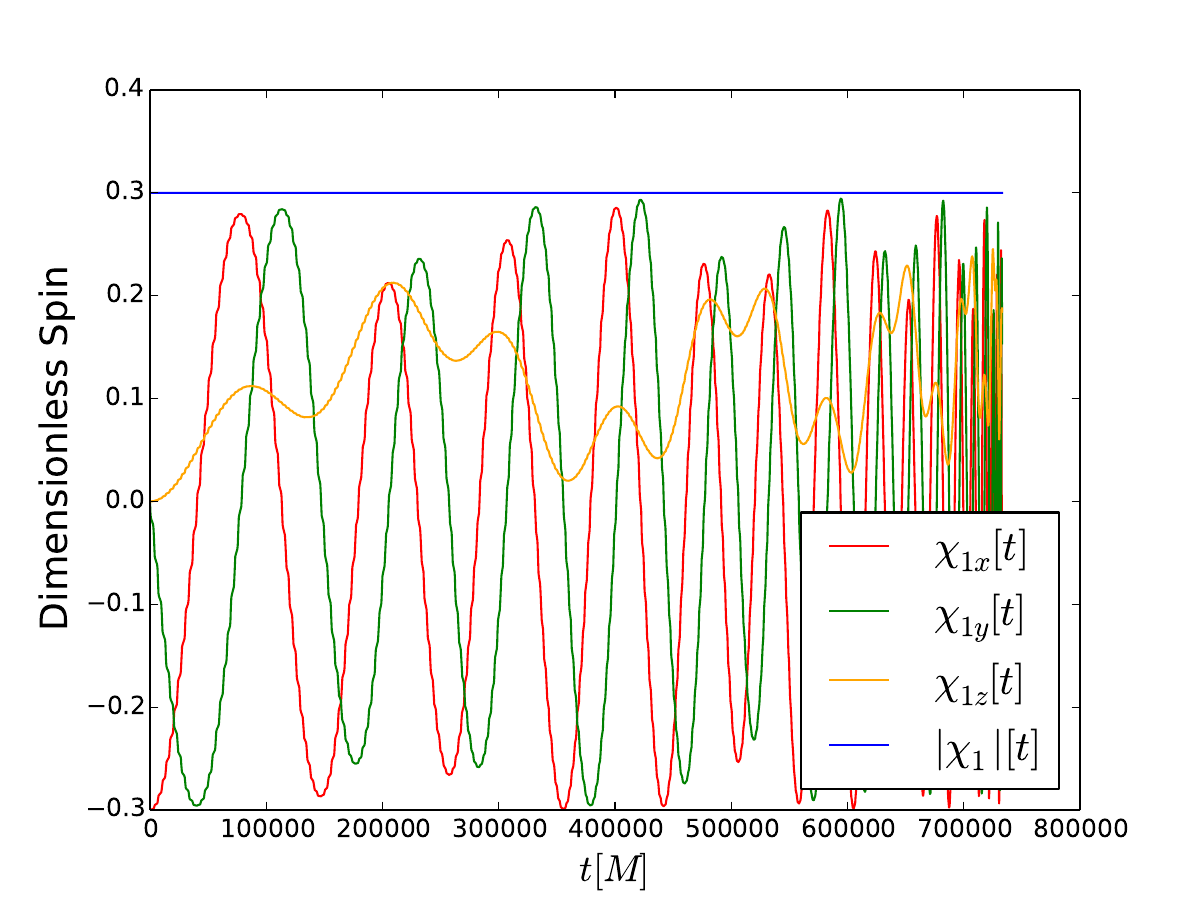}
\includegraphics[width= 0.67 \columnwidth,clip=true]{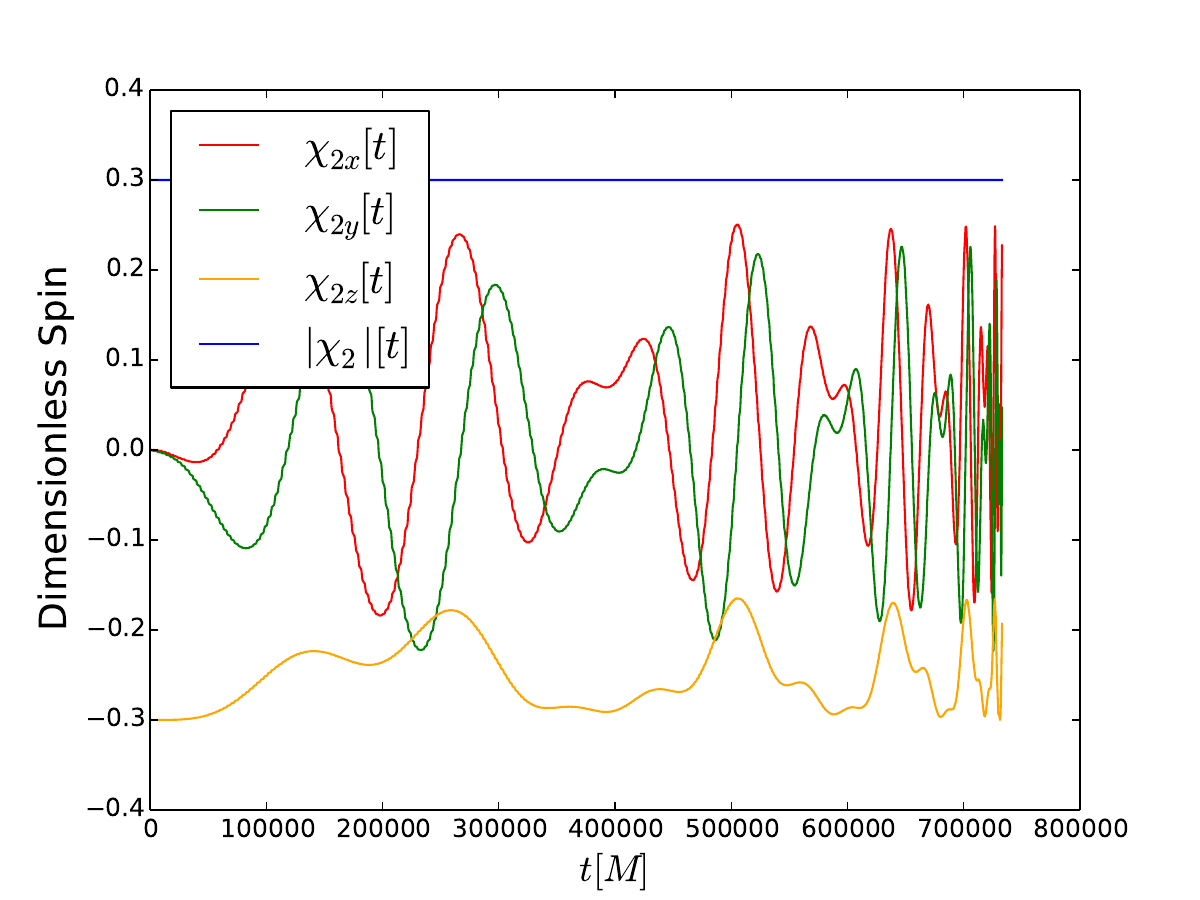}
\caption{
{\it Left}:
The individual components of the total spin vectors as a function of time
for the fiducial binary system that we outlined below.
The spins are precessing about the total angular momentum.
{\it Center}: 
The individual components of $\vec{\chi}_1$ as a function of time. 
The spin magnitude is conserved in time.
{\it Right}:
The individual components of $\vec{\chi}_2$ as a function of time. 
The spin magnitude is conserved in time.
The binary has an initial periapsis 
passage of $35M$, an initial eccentricity of $0.4$, a mass ratio $q = 3/2$, and 
initial dimensionless spin parameter values
of $\chi_1 = (-0.3, 0, 0)$ and $\chi_2 = (0, 0, 0.3)$.
}
\label{fig:Ecc_prec_spins}
\end{figure*}

We start by evolving this binary and plotting the orbital trajectories
and spin vectors, shown in Figs.~\ref{fig:sepvst}, \ref{fig:Ecc_prec_orbits},
and \ref{fig:Ecc_prec_spins}.
The eccentricity of the orbit is measured following Eq.~\eqref{geom_ecc_def},
from which we calculate the eccentricity in the beginning of the evolution as 
$e_{\rm init} = 0.449$ and the final one as $e_{\rm fin} = 0.074$.
The discrepancy between the initial eccentricity in the code
and the eccentricity that we calculate is
due to the setup of the initial conditions,
as the initial inputs to the code are calculated as Newtonian initial parameters
given a PN $\dot{r}$ and $\omega$, which still does not fully
incorporate the PN effects.
 
The eccentricity reduction can be observed fully in the orbital separation
as a function of time (see Fig.~\ref{fig:sepvst}),
and efficiently radiates the eccentricity away over the evolution.

\begin{figure*}[tb!]
\centering
\includegraphics[width= 2 \columnwidth,clip=true]{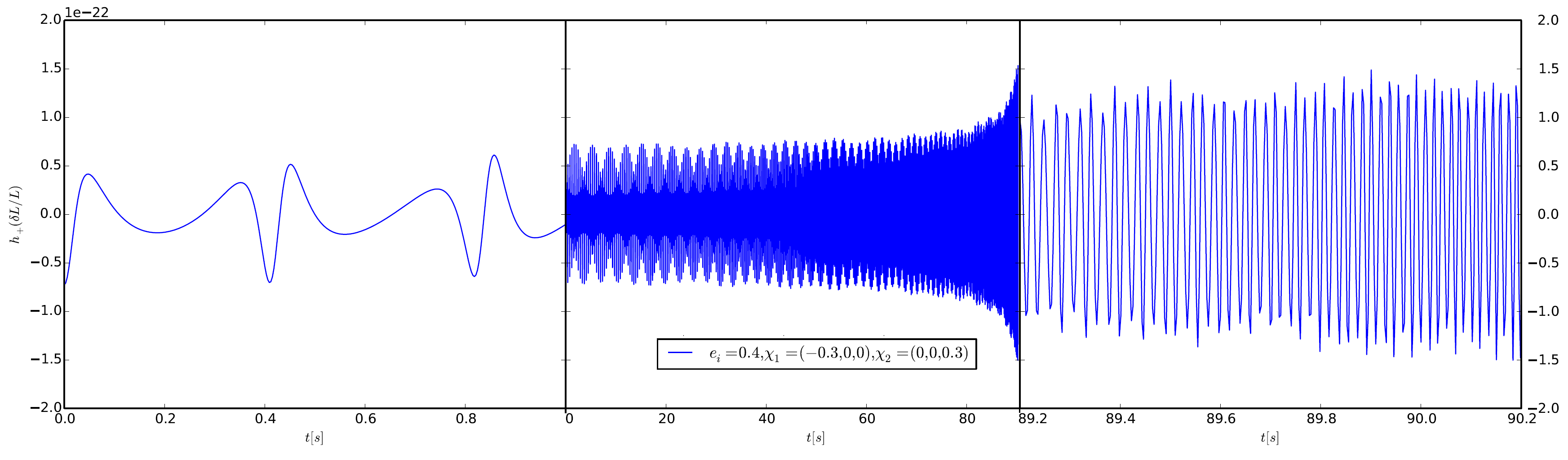}
\caption{
The plus polarization of the GW waveform starting with an eccentricity of $0.4$,
and the dimensionless spin vectors $\chi_1 = (-0.3, 0, 0)$ and $\chi_2 = (0, 0, 0.3)$.
The horizontal axis is time in seconds,
where we have dimensionalized the units by using a total binary mass
of 25 solar masses. The vertical axis is dimensionless strain $\delta L/L$,
where we have calculated the strain with a distance to the source at $500$Mpc,
and optimal orientation of the binary.
See Sec.~\ref{sec:ecc_spin_binaries} for discussion on initial conditions.
The measured initial frequencies are $3.5$\,Hz at apoapsis,
and $10.5$\,Hz at periapsis.
The measured end frequencies are $55.5$\,Hz at apoapsis,
and $66.2$\,Hz at periapsis.
The left and right panels show close-ups of the first and last seconds,
respectively.
In addition, the spin precession frequency of the GW is roughly $0.1$\,Hz.
}
\label{fig:Ecc_prec_waveform}
\end{figure*}

\begin{figure*}[htb!]
\centering
\includegraphics[width= 2 \columnwidth,clip=true]{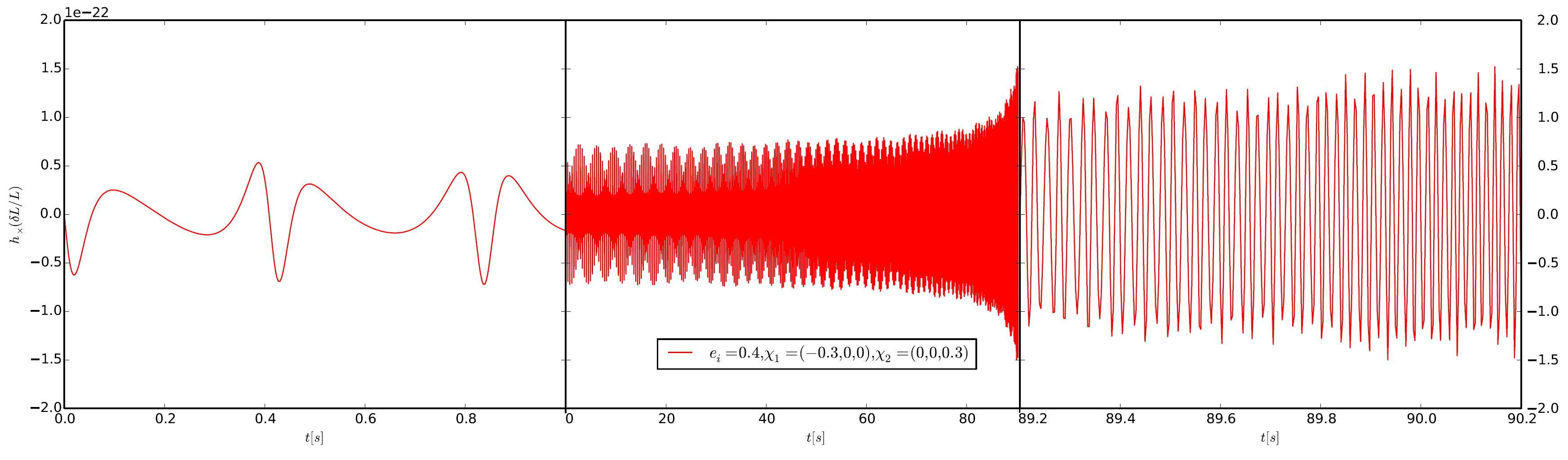}
\caption{
The cross polarization of the GW waveform with the same parameters
as the plus polarization shown in Fig. \ref{fig:Ecc_prec_waveform}.
The frequencies are the same as for the plus polarization,
but the phase is $90^{\circ}$ off (for this system and orientation).
}
\label{fig:Ecc_prec_waveform2}
\end{figure*}

With the orbital quantities, we can calculate the waveforms as well,
using the GW waveform prescription in Ref.~\cite{Will:1996zj},
using the quadrupole formula up to 2PN corrections in Eq.~\eqref{eq:2PNQPwave}. 
The plus and cross polarizations 
are tabulated in Figs.~\ref{fig:Ecc_prec_waveform}
and \ref{fig:Ecc_prec_waveform2}.

We calculate the initial and final periapsis and apoapsis GW frequencies 
directly from the waveform as opposed to an approximate GW
frequency from the orbital frequency in order to be as accurate as possible
in the calculation of the GW frequency for the entry
into a relevant detector (such as LIGO). 
For this binary, the measured initial frequencies are $3.5$\,Hz at apoapsis
and $10.5$\,Hz at periapsis,
and the end frequencies are $55.5$\,Hz at apoapsis and $66.2$\,Hz at periapsis.

The spin precession frequency is calculated in
the same way as the periapsis and apoapsis frequencies by
the peak to peak calculation on the waveform. This is 
not an exact calculation for this frequency, as the spin precession has 
some ambiguity in the waveform. Since one needs to be able to define the 
peak in the precessional modulation, this may off by an orbit or two. 
For the purposes of these calculations, we use the peak amplitude in the waveform 
for each larger spin precession modulation, getting a value of $0.1$\,Hz.

An interesting note with these calculations is that the peak periapsis frequency 
actually drops with time. 
This is an artifact of averaging over half an orbit, and when calculating the 
frequency through $\mathrm{d} \phi/\mathrm{d} t$, this is not an issue.
The period in the later stages of the orbit are cleanly broken up
into periapsis and apoapsis frequencies.
We observe that the sharp periapsis smooths out as the binary circularizes,
as expected.

\section{Comparisons and Validation}\label{sec:comp_T4_EOM}

We now turn to comparing the methodology of the direct 
integration to our implementation of the Taylor T4 method
(see Appendix~\ref{sec:T4_method}
based on, e.g., Ref.~\cite{Boyle:2007ft}),
restricted to the case when both should be valid. 

Taylor T4 is a PN approximant that assumes quasi-circular, 
non-precessing orbits. So to compare to Taylor T4, we restrict ourselves
to black holes (BHs) that are sufficiently separated,
have vanishing orbital eccentricity,
and with both spins only along the axis of the orbital angular momentum $\vec{L}$
(each could be either aligned, anti-aligned, or zero).

To begin, we will define a consistent comparison for both 
Taylor T4 and the direct integration EOMs, and 
outline the steps needed to reduce to this comparison. We will then compare the 
orbital frequency and other orbital quantities,
and the GW waveforms generated by these systems.

\subsection{Quasi-Circular Orbits in the Direct Integration}\label{sec:quasi_circular}

Until now we have made no assumptions about eccentricity; all of the
expressions to this point are valid for general orbits.
Here, we reduce the EOMs to the case of quasi-circular orbits, which 
is important for our comparisons to other quasi-circular approximants.

To do this, following Ref.~\cite{Faye:2006gx}, we introduce a moving 
orthonormal triad $\{\hat{n},\hat{\lambda},\hat{\ell}\}$, 
where $\hat{n}$ is the same as above, 
$\hat{\ell}=\hat{n}\times\vec{v}/|\hat{n}\times\vec{v}|$, and 
$\hat{\lambda}=\hat{\ell}\times\hat{n}$ (see Appendix~\ref{sec:oEOM}).
Notice that $\hat{n}$ and 
$\hat{\lambda}$ span the orbital plane, while $\hat{\ell}$ is normal to it.
In this basis, general kinematic 
considerations lead to the acceleration equation
\begin{eqnarray}
\frac{\mathrm{d}\vec{v}}{\mathrm{d}t} = (\ddot{r} - r\Omega^2)\hat{n}
+ (r\dot{\Omega} + 2\dot{r}\Omega)\hat{\lambda} + r\varpi\Omega\hat{\ell} \,,
\label{(B:390)}
\end{eqnarray}
where the dot denotes $\mathrm{d}/\mathrm{d}t$,
$\Omega$ is the orbital frequency, and $\varpi$ is the orbital plane 
precession frequency defined by $\varpi=-\hat{\lambda} \cdot \mathrm{d}\hat{\ell}/\mathrm{d}t$.
Quasi-circular orbits $\ddot{r}\simeq\dot{r}\simeq\dot{\Omega}\simeq0$, 
the acceleration equation reduces to
\begin{eqnarray}
\frac{\mathrm{d}\vec{v}}{\mathrm{d}t} \simeq -r\Omega^2\hat{n} + r\varpi\Omega\hat{\ell} \,.
\label{(B:391)}
\end{eqnarray}
By identifying Eqs.~\eqref{(B:391)} and \eqref{pEOM}, one obtains 
PN expansions for $\Omega^2$ and $\varpi$.

Introducing the frequency-related parameter 
$x\equiv(M \Omega)^{2/3}$ gives a relation $x(\gamma)$,
which may then be inverted order-by-order to obtain a relation 
$\gamma(x)$. With
$\gamma(x)$ in hand, one may re-express any function of the 
coordinate-related parameter $\gamma$ in terms of the frequency-related
parameter $x$. This is often preferable, since expressions in 
terms of the frequency-related parameter are gauge-invariant, whereas 
expressions in terms of the coordinate-related parameter are not.
Examples of quantities that are useful to express as functions of 
$x$ are the energy, $E(x)$, the flux, $\mathcal{F}(x)$, and the orbital 
phase, $\phi(x)$. Expressions for these are given in Ref.~\cite{Brown:2007jx}
in terms of phase variable $v$ (see Appendix~\ref{sec:T4_method}),
and is the jumping off point for the Taylor T4 
approximant (see, e.g., Ref.~\cite{Boyle:2007ft}).

\subsection{Eccentricity Reduction}\label{sec:ecc_reduction}

For comparing to adiabatic methods such as Taylor T4, we need to be able 
to accurately and reliably give quasi-circular initial conditions to the direct 
EOMs numerically. This is more challenging than at first glance, because if
we were to simply give Newtonian
(or even PN~\cite{Healy:2017zqj, Ramos-Buades:2018azo})
initial conditions (detailed above), there would 
be an error on the order of the neglected PN terms in the initial trajectories. 
This would manifest as a spurious eccentricity and add undesirable 
dynamics into the simulation. 

We follow a simple procedure to remove this
unwanted eccentricity. This procedure has been developed to set up low
eccentricity numerical relativity initial
data~\cite{Pfeiffer:2007CQGra..24S..59P, Buonanno:2011PhRvD..83j4034B}.
We begin by modeling the inspiral as a superposition of two effects, the (real) inspiral, which is a
smooth decrease in the orbital separation as a function of time, and the (unphysical) oscillation due to the spurious eccentricity.

\begin{figure}[!t]
\begin{center}
\includegraphics[width=\columnwidth,clip=true]{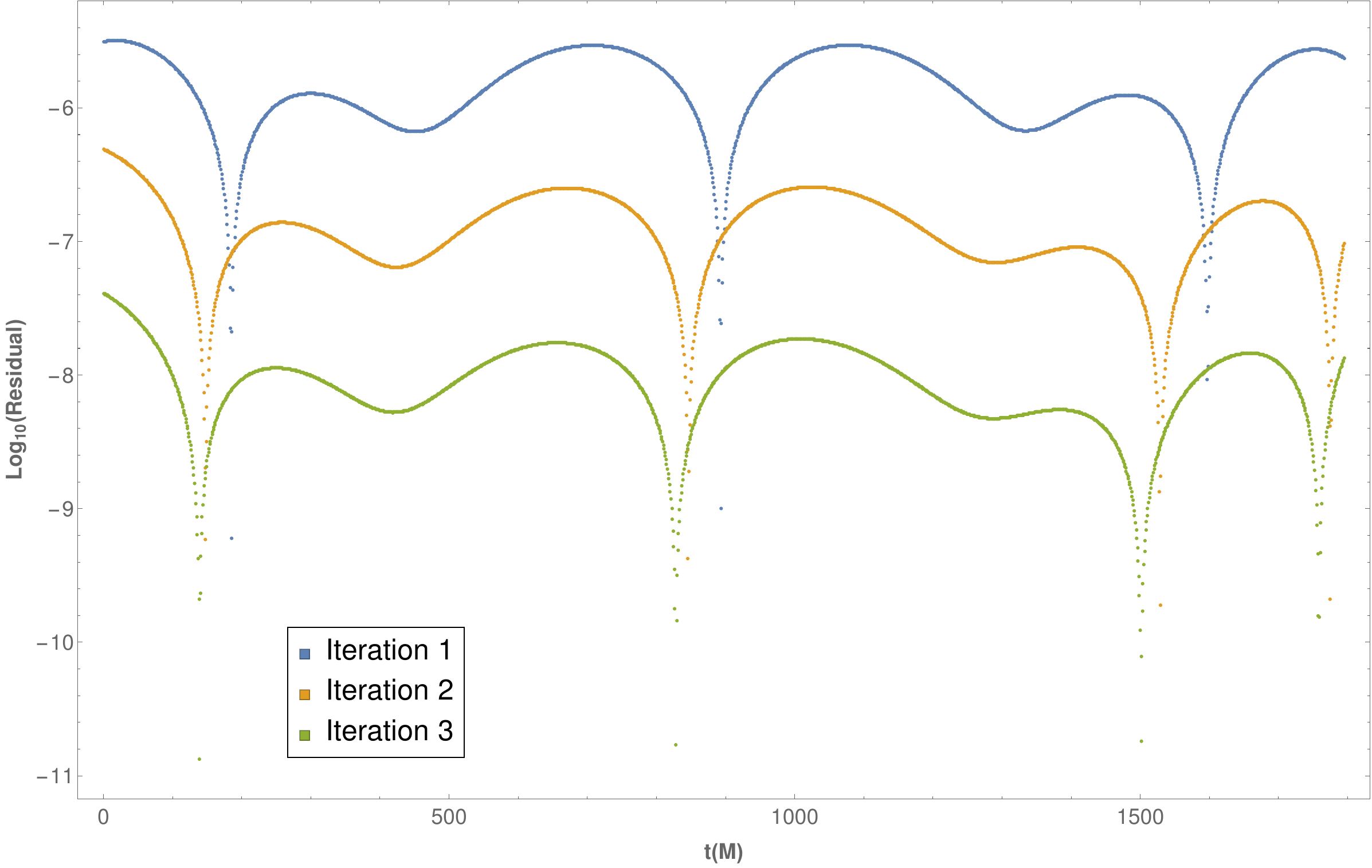}
\end{center}
\caption{
Log of the residuals of $\dot{r}$ in the eccentricity removal from the data
and the model~\eqref{rdot_model}, over the first three orbits of evolution
to refine the initial conditions.
This example is at an initial separation of $20M$, 
with equal masses, and no spin. 
}
\label{fig:Ecc_Remover_Resid}
\end{figure}

We start with a simple assumption for the inspiral part and oscillatory part,
namely:
\bea \label{rdot_model}
\frac{\mathrm{d}r}{\mathrm{d}t} = v_{\mathrm{insp}}(t) + B \cos( \omega\, t + \varphi) \,.
\eea
We take the inspiral model $v_{\mathrm{insp}}(t)$ to be a simple polynomial,
which we can fit for $v_{\mathrm{insp}}(t) = v_0 + v_1 t + v_2 t^2$
with coefficients $v_0$, $v_1$ and $v_2$, and the oscillatory 
piece $B \cos( \omega\, t + \varphi)$
with the amplitude $B$, frequency $\omega$, and initial phase $\varphi$.
With this model, we run our EOM code 
with the quasi-circular initial conditions detailed above, and fit the 
data with this model. We can then subtract out the oscillatory piece, and 
iterate on this model as many times as we need to attain the quasi-circular
initial conditions. 
This is illustrated in Fig.~\ref{fig:Ecc_Remover_Resid}.

Mathematically, this is taking the initial conditions
$\dot{r}$ and $\Omega$, and after fitting the inspiral, 
we update 
\bea
\dot{r}_{\rm new} = \dot{r}_{\rm old} + \Delta \dot{r} \,,
\eea
and 
\bea
\Omega_{\rm new} = \Omega_{\rm old} + \Delta \Omega \,,
\eea
with $\Delta \dot{r}$ and $ \Delta \Omega$ given by:
\begin{subequations}
\bea
\Delta \dot{r} &=& - B \cos \varphi \,, \\ 
\Delta \Omega &=& - \frac{B \omega \sin \varphi}{2 r_0 \Omega_0} 
\approx - \frac{B \sin \varphi}{2 r_0} \,,
\eea
\end{subequations}
where $r_0$ is the initial separation of the binary,
and $\Omega_0$ is the initial orbital frequency.

\subsection{Consistent PN Order in Taylor T4}\label{sec:consistent_T4}

The flux formula in Taylor T4 is higher order than what we can calculate in the EOM 
formalism. To combat this, we need to tailor the T4 fluxes to a consistent PN order. 
From, e.g., Eq.~(A.13) of Ref.~\cite{Brown:2007jx},
we can see that the leading order of $v$ is 
a 2.5PN term. We can also see the series expansion in the flux is out to $v^7$,
which is 3.5PN \textit{beyond} leading order.
The absolute PN order for the fluxes is then 
6PN, which is far beyond the highest order terms in the EOM formalism.
To be consistent with our non-adiabatic EOM formalism above, we 
must truncate the flux terms above next to leading order
so that the total re-expanded rational fraction is consistently 3.5PN,
i.e., for non-spinning binaries:
\bea
\mathcal{F}(v) &=& \frac{32}{5} v^{10} \eta^2 \bigg \{ 1 + v^2 \bigg( - \frac{1247}{336} - \frac{35}{12} \eta \bigg) \bigg \} \,.
\eea

This is what we mean by a ``consistent'' PN order when we compare to the EOMs
in the following sections. When quantifying comparisons,
we will use the consistent PN order and the high PN order
(keeping the flux terms to 6PN) to track the effects of PN orders.

\subsection{Orbital Frequency Comparison: No Spin}\label{sec:orbital_freq}

\begin{figure*}[t!]
\includegraphics[width= 0.66 \columnwidth,clip=true]{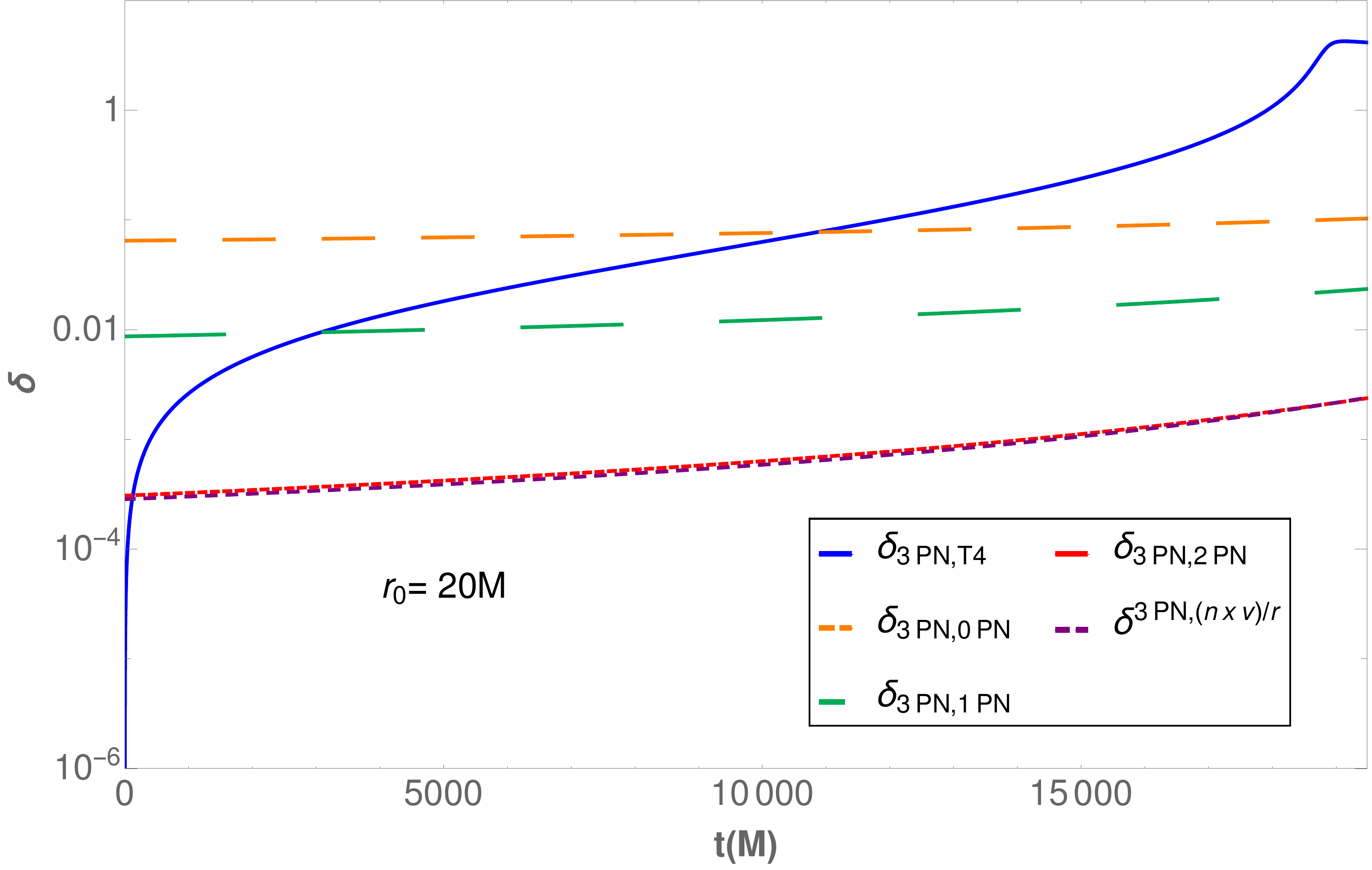}
\includegraphics[width= 0.66 \columnwidth,clip=true]{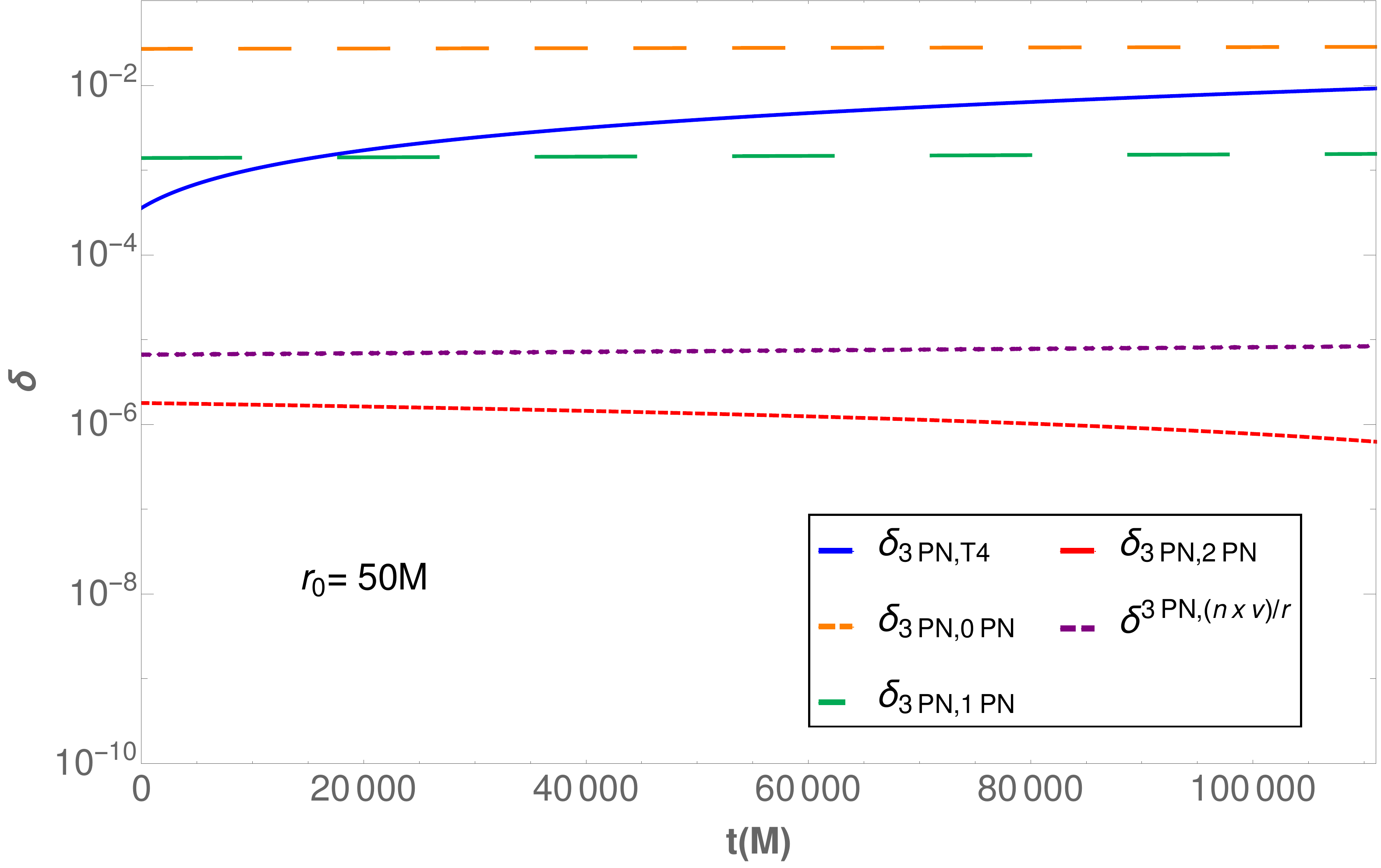}
\includegraphics[width= 0.66 \columnwidth,clip=true]{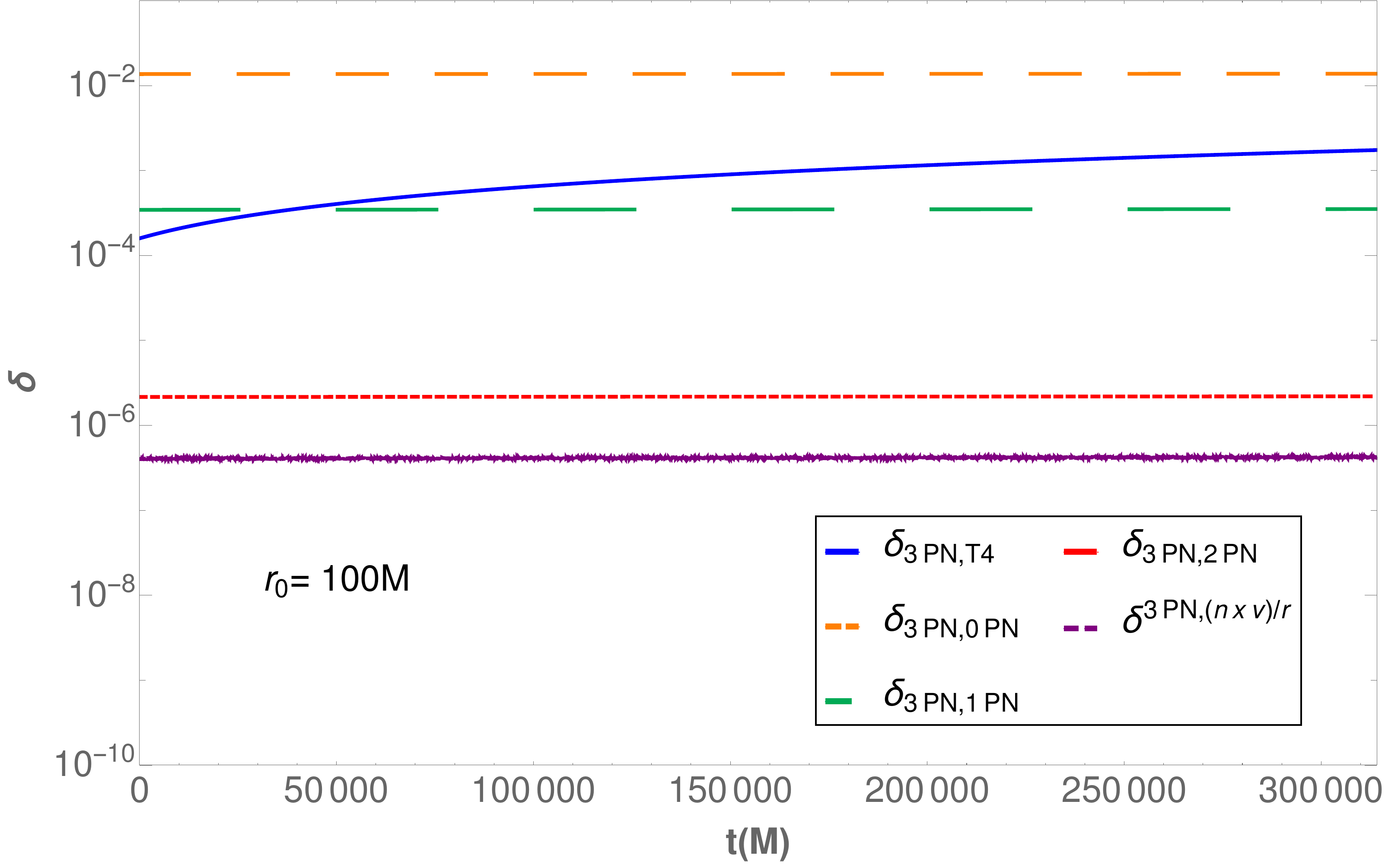}
\caption{
The relative error $\delta({\rm 3PN,Aprx2})$ of the orbital frequencies
as a function of time at different PN orders, starting the binary
at orbital separations of $r_0=20M$ (left), $r_0=50M$ (center) and $r_0=100M$ (right).
All lines compare the 3PN term in Eq.~\eqref{PN_omega_def} to other approximants.
The orange (dashed) line compares 3PN to the Newtonian (0PN) term,
so the leading order PN error is 1PN.
The green/red (longer-dashed/shortest-dashed) line compare to 1PN/2PN,
so the leading order error is 2PN/3PN.
The purple (shorter-dashed) line is the 
relative error of the different orbital frequency definitions
for the EOM code, which is similar to
the 3PN error for the entire evolution.
The blue (solid) line is the relative error between the 3PN EOM code
and the consistent PN order Taylor T4. 
}
\label{fig:Multi_M_freq_reldiff}
\end{figure*}

With the orbital quantities from solving Eq.~\eqref{T4_ev}, and the direct EOM 
orbital quantities from solving Eq.~\eqref{direct_EOM},
we can explore how to compare these to one another. 

First, we need to give both methodologies the same initial conditions, to 
ensure an apples to apples comparison.
This is achieved by using the orbital frequency calculated for the EOM orbit initially
(see Eq.~\eqref{PN_omega_def} below), and setting the initial frequency of the 
T4 code to the same initial value. This is simple using the relation 
$\Omega = (v^3/M)$, and solving for $\Omega$. We also define the binary phase 
to start in the same location on the orbit, i.e.,
the binary starts on the $x$-axis and 
the angular momentum is defined in the usual right handed sense.

The first and most direct comparisons that we can make are
with the orbital frequencies themselves.
The orbital frequency via the Taylor T4 method is given by the second 
integration equation, $\mathrm{d}\phi/\mathrm{d}t = v^3/M$,
as $\mathrm{d}\phi/\mathrm{d}t = \Omega$. 

When we talk about the direct EOM method, however, we need to be careful in 
how we define the orbital phase. 
Since EOM method directly outputs the trajectories and velocities,
we can calculate the orbital frequency in the Newtonian sense: 
\bea
\Omega = \frac{| \vec{r} \times \vec{v} |}{r^2} \,,
\eea
which specify the orbital frequency given the orbital trajectories and
velocities, $\vec{r}$ and $\vec{v}$.
To contrast this, we can use an alternate definition from PN given
in Ref.~\cite{Blanchet:2013haa} for no spins as 
\begin{widetext}
\bea \label{PN_omega_def}
\Omega^2 &=& \frac{M}{r^3} \bigg\{ 1 + \bigg( - \frac{7}{4} + {1}{4} \eta \bigg)\gamma 
+ \bigg( - \frac{7}{8} + \frac{49}{8} \eta + \frac{1}{8} \eta^2 \bigg) \gamma^2 \\ \nonumber
&& + \bigg( - \frac{235}{64} + \bigg[ \frac{46031}{2240} - \frac{123}{64} \pi^2 
+ \frac{22}{3} \ln \bigg(\frac{r}{r_0'}\bigg)\bigg] \eta + \frac{27}{32} \eta^2 + \frac{5}{64} \eta^3 
\bigg) \gamma^3 \bigg \} + O \bigg(\frac{1}{c^8}\bigg) \,,
\eea
\end{widetext}
where $r_0'$ is a gauge constant that we set to $10M$,
which is also used in the direct EOM method,
and $\gamma$ is the PN parameter $M/r$.
This allows us to quantify a PN order by dropping higher order terms
and plotting the differences. 

For the following analysis, we are restricting to a non-spinning binary
in a quasi-circular orbit, 
that starts in the same initial position for the evolutions of the binary
with both Taylor T4 and our direct EOM integration.

We give the results of these different 
PN orders in Fig.~\ref{fig:Multi_M_freq_reldiff}
by calculating
\bea
\delta_{\rm aprx1,aprx2}
= \left| \frac{\Omega_{\rm aprx1} - \Omega_{\rm aprx2}}{\Omega_{\rm aprx1}}\right|
\,.
\eea
We can see immediately that they follow a strict hierarchy of decreasing
relative difference as the PN orders are increased,
i.e., the differences $\delta_{\rm 3PN,iPN}$ decrease with increasing PN order $i$,
which is a good initial sanity check.

\begin{figure}[b!]
\includegraphics[width= \columnwidth,clip=true]{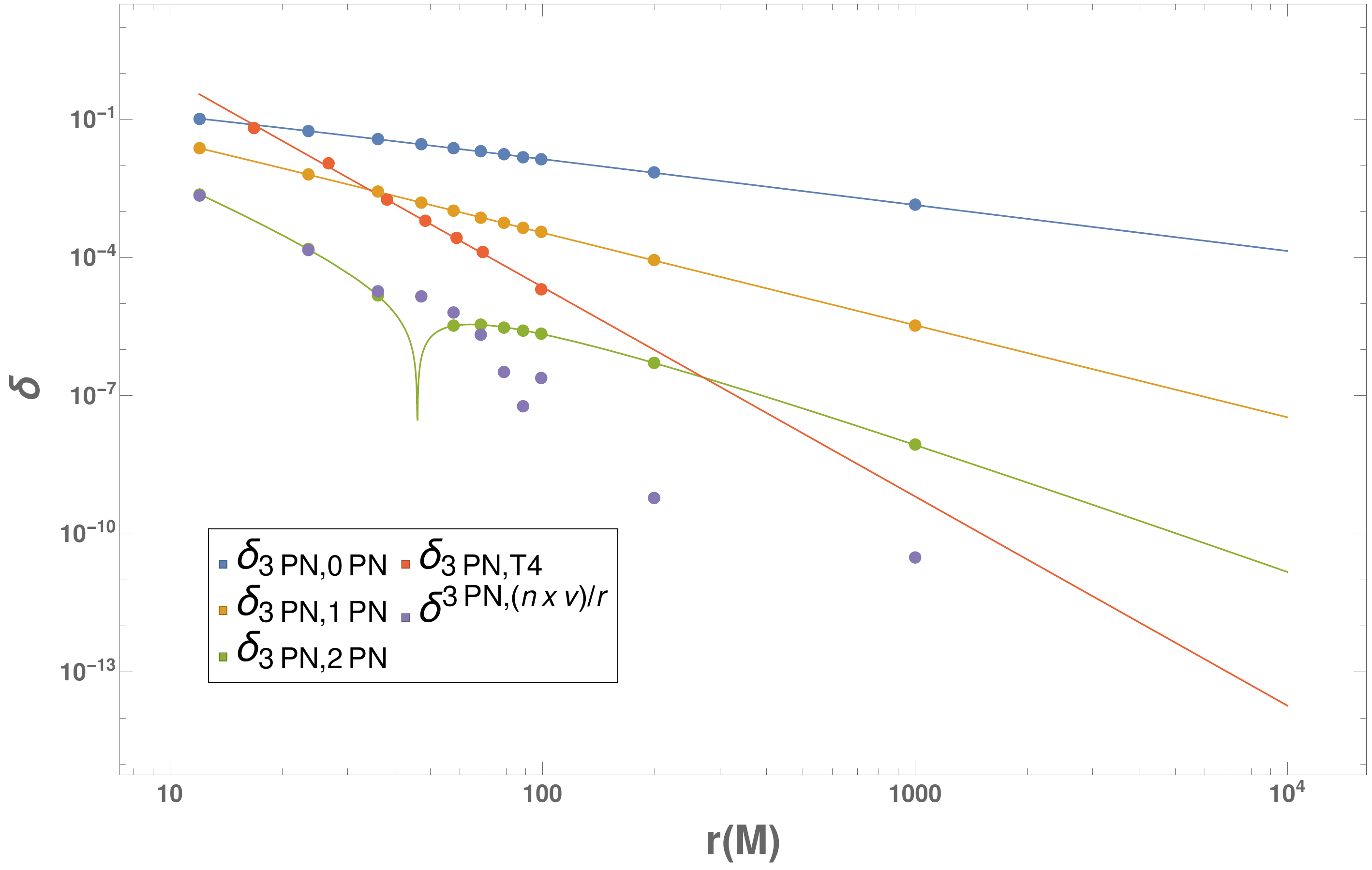}
\caption{
Tracking the PN order of the orbital frequencies as a function of time.
We measure the slope of the functional form of the orbital frequency (in solid),
through the evolution data (points) for the orbital frequency PN scalings.
The T4-EOM comparisons are obtained by fitting the data
to a simple polynomial and over plotting the resultant fit function through the data.
}
\label{fig:freq_reldiff_PNeffect}
\end{figure}

We similarly define the difference stemming from using different definitions of the frequency itself,
\bea
\delta^{\rm def1,def2}
= \left| \frac{\Omega^{\rm def1} - \Omega^{\rm def2}}{\Omega^{\rm def1}}\right|
\,,
\eea
and note that the difference $\delta^{\rm 3PN,n \times v / r}$
accounts for roughly the same order error as the 3PN term
(which is reassuring, since the orbital frequency is given to 3PN order
in the alternate PN definition stated above). 
There is a curious thing happening at the initial separation $r=50M$,
where this difference $\delta^{\rm 3PN,n \times v / r}$
is larger than the 3PN error. 
This is explained in Fig.~\ref{fig:freq_reldiff_PNeffect},
where it becomes apparent that the 3PN term has a zero crossing,
and thus dips lower than the difference from the frequency definition.
This zero crossing is directly dependent on the gauge term in
Eq.~\eqref{PN_omega_def},
and can be shifted at will for different choices of $r_0'$. 

The relative difference of separate PN orders gives a measure of the error
in terms of the PN order. We can use this to quantify any comparison
in terms of the orbital frequency PN order, and to check the PN scaling
of the discrepancies of Taylor T4 and the EOM methods,
$\delta_{\rm 3PN,T4}$.
What we find is a strong scaling as a function of separation.
When the binary is closely separated ($r\leq 12M$), 
the discrepancy is greater than $1\%$ after an initial discrepancy
below the 3PN line. As the binary inspirals,
the discrepancy grows, crossing the 2PN line after a short amount of evolution
($t\sim3000M$), and crosses the 1PN line after ($t\sim10800M$).
When we look at larger separations of $50M$, we see again this same trend
of a sharply scaling function of separation, but in this case,
it stays below the 2PN line. When we look at the $100M$ case,
we find that the discrepancy between the Taylor T4 and EOM codes again stays
below the 2PN line, but with a lower relative difference. 

Of course, to demonstrate that this is indeed a PN scaling and not another effect,
we need to show that the relative differences between the PN orders scale
as the proper powers of $r^s$ when we look over a large range of separations.
We do this in Fig.~\ref{fig:freq_reldiff_PNeffect}.
This figure shows the relative differences scaling as a function of separation.
In each case, we fit a power law to the line, and record the slope.
The slopes that we find scale with orbital separation $r^{s}$, 
where the slope $s$ is measured for the individual relative errors.
We find that the 1PN error scales with separation as $s \approx -0.976$,
the 2PN error scales with separation as $s \approx -1.9995$,
the 3PN error scales with separation as $s \approx -4.002$
at far separations ($ r \approx 1000M$),
and $s \approx -2.7577$ at small separations ($r \sim 12M$).
The 3PN error does not scale as $s \approx -3.0$ because of the gauge dependent 
logarithm term.
When looking at separations that are even farther out than $r \approx 100M$,
like $r \sim 1000000M$, we find that the slope approaches
the correct PN scaling of $-3$. 
The scaling of the different PN definitions is also tracked here, 
and while there is not a clear power law trend on this plot,
we do note that the definitional discrepancy falls at or below the 3PN line
(except for the special case around $50M$), so we can say from this 
that either definition is acceptable in the EOM code.
For our purposes, we will use the 
geometric definition of $\Omega = |\hat{n} \times \vec{v}|/r$.
When we measure the scaling of the Taylor T4--EOM relative error
in Fig.~\ref{fig:freq_reldiff_PNeffect},
we find it is $s \approx -4.545$. 
We do not attribute this discrepancy to the PN order
since this does not obey any obvious PN scaling. 

This begs the question of what is 
causing this rapid drop in relative difference as the separation is increased. 

To investigate this, we need to consider what kinds of effects could be at play:
this could be a numeric effect (e.g., one of the codes is not calculating to the 
requisite precision, which is causing a discrepancy at close separations where 
time steps are smaller); an eccentricity that is deviating
the orbit from quasi-circularity in the EOM code;
or an effect of the adiabatic approximation breaking down.

To test whether or not this is a numeric effect, we doubled the precision of both 
codes and re-ran the test at $20M$ again, with the same results. Therefore, 
we conclude that this is not a numeric issue. To test whether this is an 
eccentric effect, we ran the EOM code with different levels of the eccentricity 
remover (which should, in principle, solve the quasi-circularity problem if indeed 
it is one), and checked the relative differences. We find preliminarily that 
the residual eccentricity does scale the relative difference
of the orbital frequencies, not quite one-to-one,
and is removed by the iterative eccentricity removal procedure.

The final plausible possibility is the adiabatic approximation 
breaking down at close separations, which leads to a high discrepancy 
between the two methods. This is the most challenging possibility to 
eliminate. The best method for tracing this is to orbit average all 
of the EOM code, essentially making it adiabatic. We have done this for the 
lowest order Peters-Mathews test (0PN (Newtonian) conservative terms, with a 2.5PN 
radiation reaction term added), and find that this does not seem to affect 
the evolution at lowest order. Of course, to show that this is indeed 
unaffected in the general sense, we need to go beyond the leading order
evolutions and demonstrate this for 3.5PN.


\subsection{Waveform Comparisons for Aligned Spins and different Mass Ratios}\label{sec:Overlaps}

With the orbital frequency analysis concluded,
we now turn to calculating the GW waveform overlaps
between Taylor T4 and the EOM methods.
We pick several fiducial separations, mass ratios,
and spins (both aligned and anti-aligned, denoted by $\chi_1$ and $\chi_2$).
The initial separations for which we choose to calculate the overlaps are 
$100M$, $50M$, and $20M$, with mass ratios $q=m_2/m_1$ ($m_2 \geq m_1$)
of $1$, $2$, $10$, and $100$. 
The spin parameters that we pick are $\chi_1 = \{ 0, 0.3, 0.6, 0.9 \}$ 
and $\chi_2 = \{ 0, 0.3, 0.6, 0.9, -0.3, -0.5, -0.6, -0.9 \}$,
at separations of $50M$ and $100M$. 

We calculate the overlap and maximize over time and phase
(e.g., Ref.~\cite{Buonanno:2009PhRvD..80h4043B}),
by calculating the inner product in the frequency domain
\bea
(h_1 | h_2) = 4 \max_{t_c} \bigg| \int^{f_{\rm high}}_{f_{\rm low}}
\frac{h_1^*(f, t_c) h_2(f)}{S_n(f)} \mathrm{d}f \bigg| \,,
\eea
where $S_n$ is a noise power spectrum density of a detector~\footnote{Here,
we used LIGO's target sensitivity curve, Zero-Detuned
High-Power (v2)~\cite{LIGONoise1};
it has since been superseded by v5~\cite{LIGONoise2}.}.
The maximization over time is handled by maximizing over $t_c$,
and the phase maximization is handled by the shifting of $h_1$ in the frequency domain.
The low frequency cutoff on the integration 
is set to a reasonable frequency for a detector
(for this analysis we set the low frequency cutoff to $10$\,Hz,
which is a reasonable if a bit ambitious lower frequency bound for LIGO).
The high frequency cutoff is not set, to capture the maximum overlap 
of the waveform if the endpoint is not exactly set to the same frequency. 

We then normalize (using the euclidean norm) over $h_1$ and $h_2$ 
to obtain the overlap:
\bea
O = \frac{(h_1 | h_2)}{\mathrm{Norm}(h_1) \mathrm{Norm}(h_2)} \,.
\eea

\begin{figure}[b!]
\includegraphics[width= \columnwidth,clip=true]{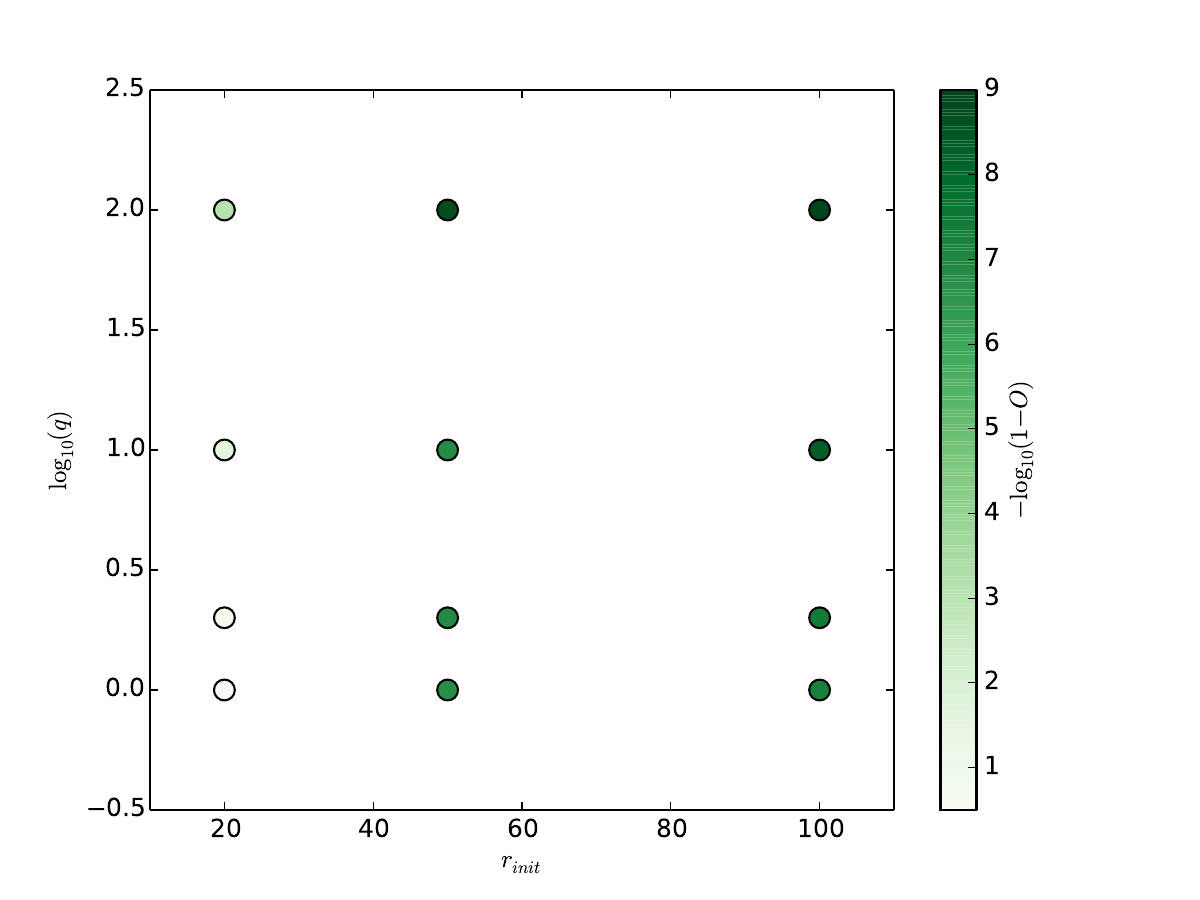}
\caption{
A visualization of the data in Table~\ref{tab:T4_EOM_overlap_comparisons},
where we have suppressed all of the spin overlaps,
and plot the initial separation vs. the log of the mass ratio,
with the color scale indicating $- \log_{10} (1 - O)$.
This overlap was done over the shortest evolution,
at 20M the simulation ran for only 20 orbits,
so the rest of the overlaps were calculated for 20 orbits
to give an accurate comparison.
This parameterization of the color scale leads to 
the darker colors indicating a better overlap
(the number of nines is indicated on the scale).
}
\label{T4_EOM_Overlap_initr_massrat}
\end{figure}

\begin{figure}[tb!]
\includegraphics[width= \columnwidth,clip=true]{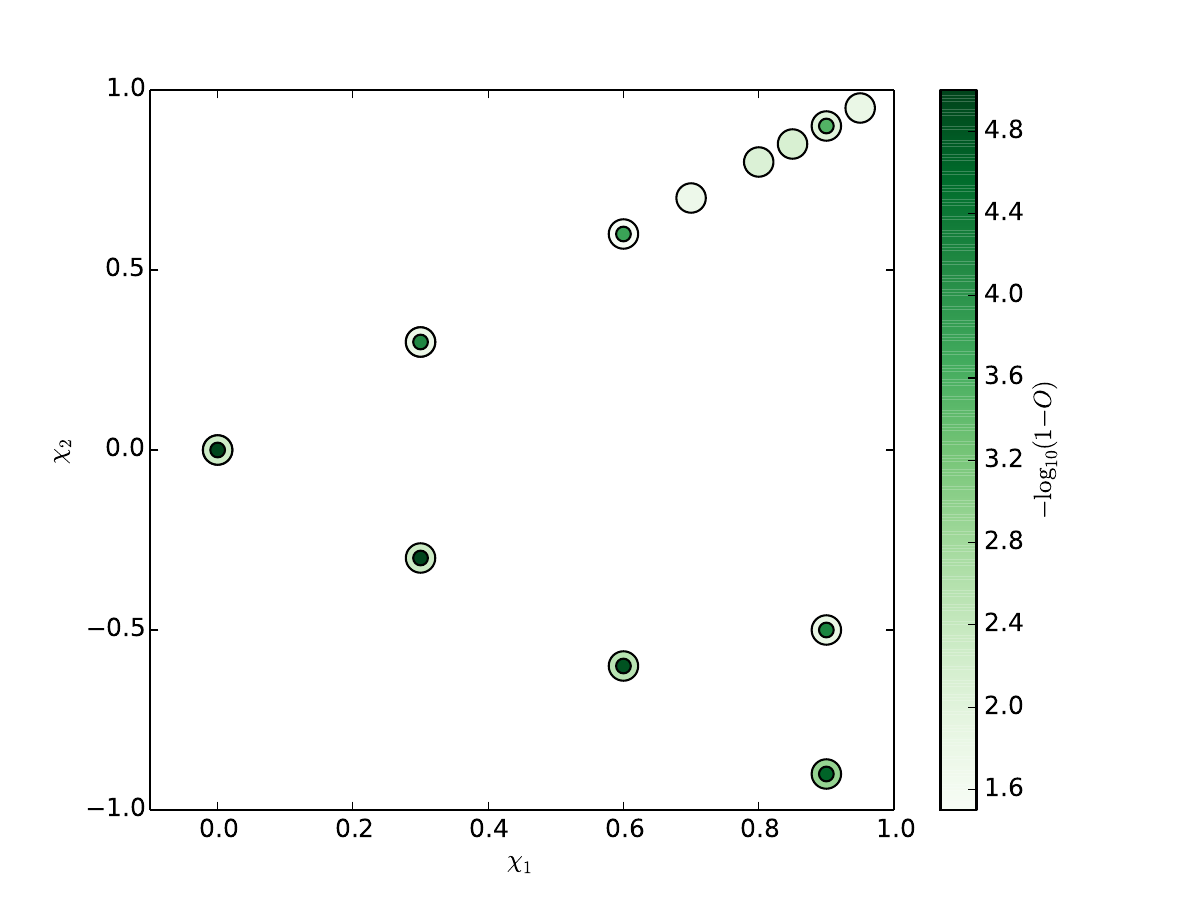}
\caption{
A visualization of the data in Table~\ref{tab:T4_EOM_overlap_comparisons},
where we have suppressed the mass ratios and initial separations,
and plot the $\chi_1$--$\chi_2$ plane,
with the color scale indicating $- \log_{10} (1 - O)$.
The large circles indicate an initial separation of $50M$,
while some configurations were also examined at $100M$,
marked with a smaller inner dot.
These overlaps were run on waveforms with a simulation duration of 50 orbits.
This parameterization of the color scale leads
to the darker colors indicating a better overlap
(the number of nines is indicated on the scale).
}
\label{T4_EOM_Overlap_chi1_chi2}
\end{figure}

The results are tabulated in Table~\ref{tab:T4_EOM_overlap_comparisons}
(see also Figs.~\ref{T4_EOM_Overlap_initr_massrat}
and \ref{T4_EOM_Overlap_chi1_chi2}).
We keep all of the parameters that we used to calculate the overlaps in the table:
the initial and final separations, the mass ratio,
the aligned dimensionless spin values $\chi_1$ and $\chi_2$,
the total simulation time in units of $M$,
the number of orbits the waveform spanned, the time step of the overlap calculation, and finally the maximized overlap for both the Taylor T4 to EOM comparison
at a consistent PN order (3.5PN), and also at the highest T4 order (6PN).

We see that the overlap is a strong function of the orbital separation:
as the separation increases, the overlap increases from a bad overlap
at $20M$ of only $O \sim 0.8$ to an overlap of $O \sim 0.99999$
at $100M$. In addition, as the mass ratio increases,
the overlap also increases. For example, at a separation of $50M$, 
holding the spins of the individual BHs to zero,
we increase in overlap from $O \sim 0.98$ at $q=1$ 
to $O \sim 0.99999$ at $q=100$.

When we move to explore the spin parameter space, we hold the mass ratio fixed
and compute the overlaps for separations of $100M$ and $50M$. 
As the dimensionless spin parameter increases
in value to higher positive $\chi$ effective
($\chi$ effective is the spin values projected along the orbital angular momentum),
the overlap goes down, at $50M$, from $O \sim 0.995$ at spins of zero,
to $O \sim 0.99$ at a high positive $\chi$ effective.

When the spins are anti aligned with each other, keeping the effective spin zero,
the overlap stays fairly constant, but increases slightly,
with the $\chi_1=0.9, \, \chi_2 = -0.9$
having an overlap of $O \sim 0.999$. 

A final discussion point is to highlight the effects of PN at higher order
with Taylor T4, specifically when we add higher order radiation reaction flux terms.
The high order T4 generically differs in overlap from the consistent order T4
by $10^{-8}$ at $100M$, $10^{-5}$ at $50M$, and $10^{-3}$ at $20M$.
Though the overlap difference increases as the separation decreases,
the PN effects at 4PN do not account for the discrepancy of the overlaps
between Taylor T4 and the direct integration EOM.

In addition to the results we tabulate in Table~\ref{tab:T4_EOM_overlap_comparisons},
we perform stability tests on the overlap between Taylor T4
and the direct integration EOM
by inputting a small $x$-component perturbation
to the two spin vectors, which will cause a small amount of spin precession,
hence marked ${\rm EOM_P}$.
Specifically, we give the dimensionless spin vectors $\chi_1 = (10^{-4}, 0, 0.3)$, 
and $\chi_2 = (-10^{-4}, 0, 0.3)$, and we run the same overlaps
with the consistent T4 method, and also run the overlap with the EOM code
with no $x$-component perturbation to the spins. 
The overlaps that we obtain are $O_{{\rm T4}-{\rm EOM}} = 0.983693663044$,
which is exactly the overlap that we obtain when running
without the $x$ perturbation.
This is easily verifiable by redoing the overlap analysis 
with the EOM code with and without the perturbation.
We obtain an overlap of $O_{{\rm EOM}-{\rm EOM_P}} = 1.0$, 
which clearly shows that a perturbation to the spin directions
do not affect the overlaps.

\section{Generic Testbed for Other Approximants: Eccentricity Definitions}\label{sec:generictest}

This direct EOM that we present is generic and can be applied broadly
to many different physical systems, and can be used as a benchmark test
for other formalisms. To illustrate this, we apply to a simple test
of orbital eccentricity put forward recently~\cite{Loutrel:2018arXiv180109009L}.


Eccentricity in general relativity is difficult and unintuitive to
define~\cite{Loutrel:2018arXiv180109009L, Chandrasekhar:1992mtbh.book.....C,
Yunes:2009PhRvD..80h4001Y, Memmesheimer:2004PhRvD..70j4011M}.
As such, most Newtonian definitions are not sufficient,
and can give wildly different results. 
To illustrate this, we are going to take two different parameterizations of 
the Newtonian eccentricity, the Runge-Lenz vector and a low eccentricity 
definition of $e$ used in NR. 

\begin{figure}[!htb]
\begin{center}
\includegraphics[width=\columnwidth,clip=true]{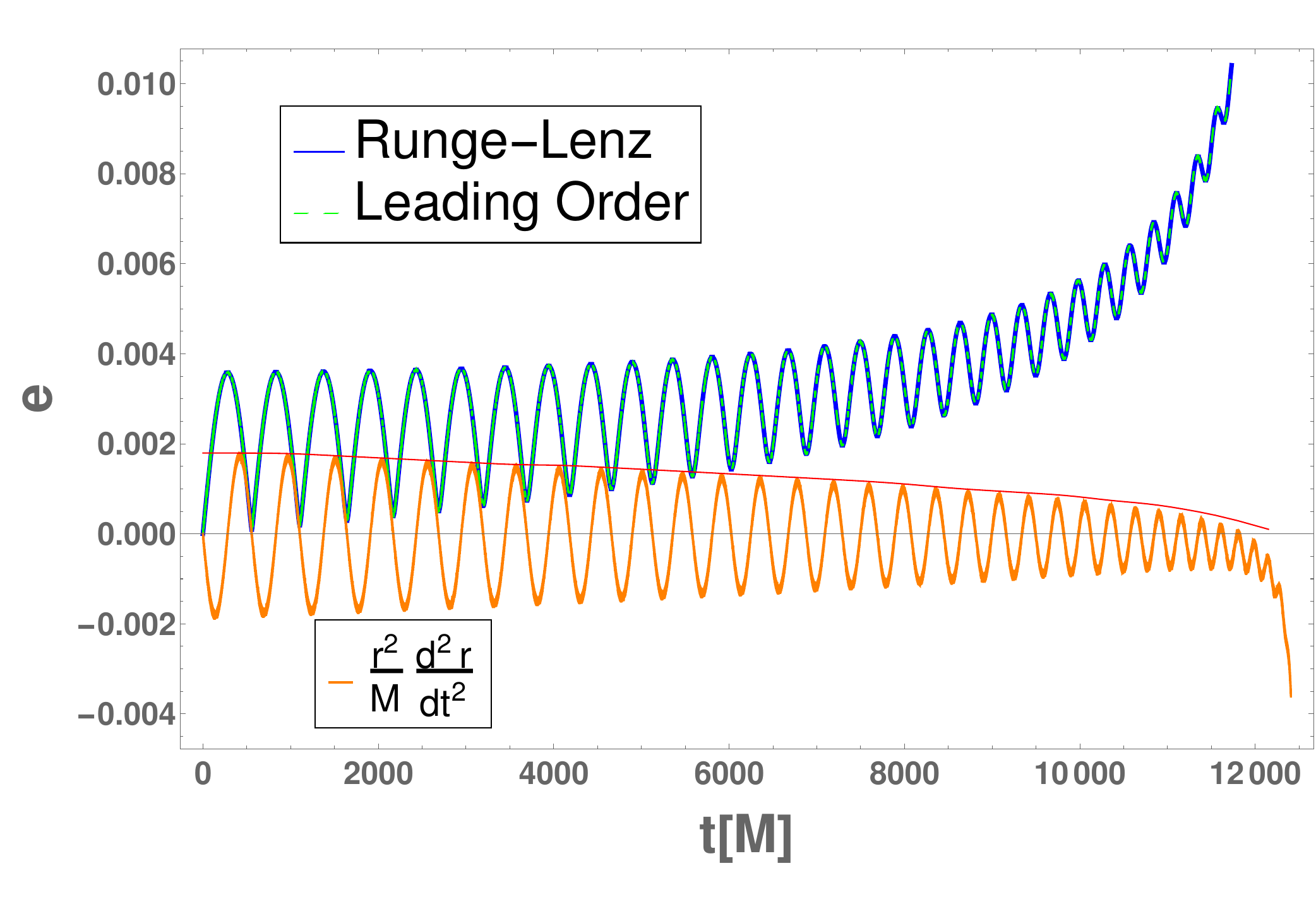}
\end{center}
\caption{
The difference that an eccentricity definition can make
in the calculated eccentricity at a given separation. 
For this figure, the BHs are non-spinning, start at $20M$ separation,
and are evolved using the orbital EOM at low order 
that contains only the 0PN (Newtonian) $+$ 2.5PN (leading-order
radiation reaction) terms.
The eccentricities are then calculated using the Runge-Lenz definition
Eq.~\eqref{RL_ecc_def} (blue),
the full NR-derived leading-order measure Eq. ~\eqref{NR_ecc_def_full} (green),
the $\ddot{r}$ formula Eq.~\eqref{NR_ecc_def} (orange),
and the amplitude of Eq.~\eqref{NR_ecc_def} (red).
The blue and green curves can be seen to be almost identical.
}
\label{fig:Ecc_Def_NR_RL}
\end{figure}

The Newtonian Runge-Lenz vector is defined as~\cite{Poisson:2014grav.book.....P}:
\bea \label{RL_ecc_def}
\vec{e} = \bigg( \frac{v^2}{M} - \frac{1}{r} \bigg) \vec{r} - \frac{\vec{r} \cdot \vec{v}}{M} \vec{v} \,,
\eea
which, for a Newtonian orbit, is a constant of the motion,
and implies that the eccentricity
itself is another constant of the motion.
To obtain the scalar eccentricity, we simply take the magnitude of this vector.
This measure is shown in Fig.~\ref{fig:Ecc_Def_NR_RL},
in a non-spinning binary system at a low PN order
at an initial separation of $20M$, as it evolves down to $10M$.

To contrast this, we will use another definition of eccentricity
taken from numerical relativity initial data,
using the radial acceleration $\ddot{r}$, 
following Refs.~\cite{Campanelli:2009PhRvD..79h4010C}
and~\cite{Husa:2008PhRvD..77d4037H}.
Under the double assumptions of low eccentricity initial data
and adiabatic frequency $\Omega$,
we can estimate $r(t)$ and its derivatives (to leading order in $e$) by
\bea
r(t) 		&\sim& M^{1/3} \Omega^{-2/3} \left( 1 + e\sin(\Omega t) \right)~, \\
\dot{r}(t)	&\sim&M^{1/3} \Omega^{+1/3} e \cos(\Omega t) ~, \\
\ddot{r}(t)	&\sim& -M^{1/3} \Omega^{+4/3} e \sin(\Omega t)~,
\eea
and then define
\bea
A(t) &=& \frac{r^2 \ddot{r}}{M}  \sim - e \sin(\Omega t)~,
\eea
so that the eccentricity is given by
\bea \label{NR_ecc_def}
e = \mathrm{Amp}(A) = \mathrm{Amp}\left(\frac{r^2 \ddot{r}}{M} \right) \,.
\eea
$A(t)$ is also shown in Fig.~\ref{fig:Ecc_Def_NR_RL}.
We note that both measures, as shown in Fig.~\ref{fig:Ecc_Def_NR_RL},
oscillate -- and for $A(t)$,
we expect the amplitude of these oscillations to be the actual eccentricity.
We thus try to complement $A(t)\sim \sin(\Omega t)$ with the corresponding
$\cos(\Omega t)$, using
\bea
B(t) &\doteq& \frac{r \dot{r}^2}{M}  \sim e^2 \cos^2(\Omega t)~,
\eea
and then should have
\bea
\sqrt{A^2 + B} &=& \sqrt{e^2\sin^2(\Omega t) + e^2 \cos^2(\Omega t)} = e ~.
\label{NR_ecc_def_full}
\eea

However, in Fig.~\ref{fig:Ecc_Def_NR_RL} we see that
\begin{enumerate}
	\item The curve for $A(t)$ drifts downwards, and at late times oscillates entirely below 0.
	\item The curve for $\sqrt{A^2 + B}$ coincides (to $\sim1\%$) with the Runge-Lenz eccentricity measure.
	\item The curve for $\sqrt{A^2 + B}$, which is supposed to serve as the non-oscillatory envelope of $A(t)$,
			does oscillate just like $A(t)$; furthermore, it both starts off above and drifts upwards faster than $A(t)$.
	\item The amplitudes of oscillations themselves shrink for both curves (which is what we would have expected the full eccentricity to do).
\end{enumerate}
As all ``physical'' eccentricities remain very low ($e\sim10^{-3}$),
this suggests that the adiabatic approximation is breaking down;
we shall analyze just how:

We no longer take $\Omega$ as a constant, but rather as $\Omega(t)$;
we model its leading order behavior as
$\Omega(t) \propto \left(t_c \!-\! t\right)^{-3/8}$,
with $t_c$ the projected leading order time of coalescence~\cite{BasicPhys150914}.
Then $\dot{\Omega}\sim\Omega/\left(t_c \!-\! t\right)$
and $\ddot{\Omega}\sim\Omega/\left(t_c \!-\! t\right)^2$.
We note that for the evolution shown in Fig.~\ref{fig:Ecc_Def_NR_RL},
$\left(t_c \!-\! t\right)$ decreases from $\sim10^4$ to $\sim10^3$,
while $\Omega$ changes over a few $\sim10^{-2}$. 

We then see that the derivatives of 
\bea
r(t) 		&\sim& M^{1/3} [\Omega(t)]^{-2/3} \left( 1 + e\sin(\Omega t) \right)
\eea
more accurately behave like
\bea
\dot{r}(t)	&\sim&M^{1/3} \Omega^{+1/3} \left[ e \cos(\Omega t) + \alpha\frac{\dot{\Omega}}{\Omega^2} + \cdots \right] ~, \\
\ddot{r}(t)	&\sim& -M^{1/3} \Omega^{+4/3} \left[ e \sin(\Omega t) + \beta\frac{\ddot{\Omega}}{\Omega^3}  + \cdots \right]~,
\eea
with numerical factors of order unity, $\alpha$ and $\beta$.

For $\ddot{r}$,
when $\left(t_c \!-\! t\right)\sim10^4$
the drift term is $\sim \left(t_c \!-\! t\right)^{-2}\Omega^{-2}
\sim(10^4)^{-2}(10^{-2})^{-2}=10^{-4}$,
and so is smaller than the oscillating eccentricity term $e$;
however when $\left(t_c \!-\! t\right)\sim10^3$,
the drift term grows to $\sim(10^3)^{-2}(10^{-2})^{-2}=10^{-2}$,
and so at late times dominates over the oscillations.
This explains why the term $A(t)$ oscillates
with the amplitude of the eccentricity early on,
but is later dominated by the drift.

For $\dot{r}$ the situation is different: even at early times,
$\left(t_c \!-\! t\right)\sim10^4$,
the drift term, $\sim \left(t_c \!-\! t\right)^{-1}\Omega^{-1}
\sim(10^4)^{-1}(10^{-2})^{-1}=10^{-2}$, is already larger than the eccentricity $e$,
and so the behavior of $B(t)$ is governed by the drift,
rather than the eccentricity oscillations.

As the Runge-Lenz definition Eq.~\eqref{RL_ecc_def} is essentially given
by the velocity components,
it adopts these features of $\dot{r}$, and measures the drift associated
to non-adiabaticity,
rather than the eccentricity itself; the eccentricity is given only by the oscillations.

\section{Discussion}\label{sec:discussion}

We have developed and constructed a direct integration of the PN EOM
in the harmonic gauge for the construction of eccentric BBHs
with arbitrary spins. This work represents a step forward for the modeling of these 
systems by extending known methods (such as Taylor T4, and the 
current LIGO/VIRGO collaboration methods) 
to binaries that evolve the spin procession 
equations and orbital motion equations for a fully generic GW waveform. 
This formalism is not limited by eccentricity as
the post-Keplerian eccentricity expansion models 
(e.g., Refs.~\cite{Gopakumar:2002PhRvD..65h4011G,Yunes:2009PhRvD..80h4001Y,Huerta:2014PhRvD..90h4016H,
Tanay:2016PhRvD..93f4031T, Hinder:2017sxy}), and is capable of 
handling eccentricity along with spin precession.

We test the validity of this method by comparing to known results from the Taylor T4 
method. In particular, we look at the orbital frequencies produced by systems 
with the same initial conditions, and compare. The results we find is that 
the relative difference in the orbital frequencies is a strong function of separation,
that does not scale with a high order PN effect, and scales 
as a function of separation as $r^s$ with $s \sim -4.5$.

This effect is not overly worrying, even though we cannot ascribe 
exactly the cause of this scaling. This 
is most likely the fundamental difference between 
Taylor T4 and our method. The various PN dynamics methods do not produce 
the same results even among the various Taylor T approximants~\cite{Buonanno:2009PhRvD..80h4043B}.
We rule out a numeric issue, and
an eccentric effect as the sole cause. We have confirmed that the adiabatic 
approximation does not affect the evolution of the binary at low order, and 
are planning on verifying this for the full 3.5PN dynamics in future work. 

This begs the question of what is 
causing this rapid drop in relative difference as the separation is increased. 

To investigate this, we need to consider what kinds of effects could be at play:
this could be a numeric effect (e.g., one of the codes is not calculating to the 
requisite precision, which is causing a discrepancy at close separations where 
time steps are smaller); an eccentricity that is deviating
the orbit from quasi-circularity in the EOM code;
or an effect of the adiabatic approximation breaking down.

To test whether or not this is a numeric effect, we doubled the precision of both 
codes and re-ran the test at $20M$ again, with the same results. Therefore, 
we conclude that this is not a numeric issue. This is an 
eccentric effect, we ran the EOM code with different levels of the eccentricity 
remover (which should, in principle, solve the quasi-circularity problem if indeed 
it is one), and checked the relative differences. We find preliminarily that 
the residual eccentricity does scale the relative difference
of the orbital frequencies, not quite one-to-one,
and is removed by the iterative eccentricity removal procedure.

The final plausible possibility is the adiabatic approximation 
breaking down at close separations, which leads to a high discrepancy 
between the two methods. This is the most challenging possibility to 
eliminate. The best method for tracing this is to orbit average all 
of the EOM code, essentially making it adiabatic. We have done this for the 
lowest order Peters-Mathews test (0PN conservative terms, with a 2.5PN 
radiation reaction term added), and find that this does not seem to affect 
the evolution at lowest order. Of course, to show that this is indeed 
unaffected in the general sense, we need to go beyond the leading order
evolutions and demonstrate this for 3.5PN.

We also compare the GW waveforms of Taylor T4 to the EOM methods, maximizing 
over time and phase, to give a quantification of the overlap. The 
results that we find are consistent with the findings of the orbital frequency
analysis: the overlap is a strong function of the orbital separation from 
$20M$ to $100M$. The 
overlaps increase strongly as a function of mass ratio, with the 
best overlaps of the waveforms we ran being at a mass ratio $q=100$.
In addition, we performed spinning waveform overlaps to test the validity of the 
spins in the EOM code. We find that the spins do not modify the overlap of the 
waveform when the effective spin of the binary remains zero, and drops the 
overlap by about a percent when the spins are aligned and low spins 
($\chi_1 = \chi_2 = 0.3$), to about two percent when the spins are moderately 
spinning ($\chi_1 = \chi_2 = 0.6$). The overlap then gets better as the spin 
gets stronger. We also tested the stability of the code to small perturbations 
in spin by giving the EOM code a small non-zero spin unaligned with the 
orbital angular momentum, and found no effect on the overlap.


Finally, we elucidated a fiducial binary system, with realistic parameters drawn 
from galactic binary simulations, to demonstrate the flexibility and power of this 
direct integration EOM code. We used initial conditions such that the binary 
would be in the very low frequency end of the LIGO band at periapsis, and 
let the binary evolve for 200 orbits. We then output the binary orbital 
trajectories, spins, and velocities. We recover the orbital plane precession 
of the binary due to spin-orbit coupling, the spin-spin precession of the individual 
spins and spin totals, the eccentricity reduction in the binary, and calculate the 
initial and final eccentricity of the binary using our geometric definition. 
We take the orbital quantities and calculate the waveform, recovering the 
eccentric signal imprinted on the outgoing gravitational radiation, estimate the 
periapsis and apoapsis frequencies of this radiation, and show the 
spin precession modulation that is also imparted on the binary. We do this 
for an optimally oriented binary, but leave the code generic so that any 
binary orientation can be used, providing us with a fully generic, precessing, 
eccentric binary GW waveform.

We hope to apply this EOM code to parameter estimation for future gravitational wave 
detections. We currently have the code implemented in Mathematica, and the analysis takes 
approximately a minute to complete. This is far too long for parameter estimation purposes currently,
and we are planning on focusing on this in the future.

There are several relevant detectors for this fiducial source.
The periapsis frequency passage is at the threshold of detectability for LIGO/VIRGO
at design sensitivity \cite{Moore:CQG2015,Martynov:2016PhRvD..93k2004M,LIGONoise2};
it will fall into the band for future LIGO upgrades such as 
A+; and third generation ground-based detectors will have
both the periapsis and apoapsis frequencies in band. 

The precessional frequency will be  
detectable by planned space based detectors such as
the LISA mission~\cite{Audley:2017drz},
though the source outlined above will be too weak for detection,
these frequencies are in band if a nearby binary happens to have these parameters.

We also note that there is 
a planned Chinese space-borne GW detector
in the millihertz frequencies, TianQin~\cite{Luo:2015ght},
and a planned Japanese space-born detector,
DECihertz laser Interferometer Gravitational wave Observatory
(DECIGO/B-DECIGO)~\cite{Seto:2001qf,Sato:2017dkf}
in between the frequency ranges of ground based and LISA detectors.
These will make it possible to detect a binary's dynamics
by multiband GW astronomy.

\section*{Acknowledgments}

B.~I. and M.~C. received support from National Science Foundation (NSF) grants 
AST-1028087, AST-1516150, PHY-1305730, and PHY-1707946.
O.~B. is supported through the Frontiers in Gravitational Wave Astrophysics Initiative 
at the RIT's Center for Computational Relativity and Gravitation.
He also acknowledges support from NSF grant PHY-1607520.
H.N. acknowledges support from JSPS KAKENHI Grant No.\ JP16K05347 and No.\ JP17H06358.
Computational resources were provided by the BlueSky Cluster at Rochester Institute of Technology.  
The BlueSky cluster was supported by NSF grants AST-1028087, PHY-0722703 and PHY-1229173.

\appendix

\section{Another expression for PN EOM}\label{sec:oEOM}

Equation~\eqref{direct_EOM} can be reformulated into the alternate form:
\begin{eqnarray}
\frac{\mathrm{d}\vec{v}}{\mathrm{d}t} &= 
-\displaystyle{\frac{M}{r^2}} \bigg[ (1+\tilde{\mathcal{A}})\hat{n} 
+ \tilde{\mathcal{B}}\vec{v} + \tilde{\mathcal{C}}\hat{\ell} \,\,\bigg]\,.
\label{pEOM}
\end{eqnarray}
We note that while the unit vector $\hat{\ell}$ is orthogonal to 
$\hat{n}$ and $\vec{v}$, the triad $\{\hat{n}, \vec{v}, \hat{\ell}\}$ does 
{\it not} form an orthogonal basis, because $\hat{n}$ and $\vec{v}$ will, in 
general, {\it not} be orthogonal to one another (although they will be 
{\it approximately} orthogonal in the special case of quasi-circular orbits). 
This equation resembles Eq.~(129) in Ref.~\cite{Blanchet:2013haa}, except for the additional
$\tilde{\mathcal{C}}$ coefficient.  
It is noted that the term
with coefficient $\tilde{\mathcal{C}}$ represents
a component of the acceleration directed out of the instantaneous orbital plane
spanned by $\hat{n}$ and $\vec{v}$. It therefore gives rise to precession of
the orbital plane and correspondingly ought to vanish identically when spins 
are aligned or anti-aligned with the orbital angular momentum.

To obtain Eq.~\eqref{pEOM}, 
we may expand each cross-product in Eq.~\eqref{direct_EOM} by using
\begin{eqnarray}
\hat{n}\times\vec{S} 
&=& \dot{r}\tilde{S}_\ell\,\hat{n} 
- \tilde{S}_\ell\,\vec{v} 
+ r\Omega\tilde{S}_\lambda\,\hat{\ell} \,,
\label{n-cross-S}
\\
\hat{n}\times\vec{\Sigma} 
&=& \dot{r}\tilde{\Sigma}_\ell\,\hat{n} 
- \tilde{\Sigma}_\ell\,\vec{v} 
+ r\Omega\tilde{\Sigma}_\lambda\,\hat{\ell} \,,
\label{n-cross-Sigma}
\\
\vec{S}\times\vec{v} 
&=& -v^2\tilde{S}_\ell\,\hat{n}
+ \dot{r}\tilde{S}_\ell\,\vec{v}
+ r\Omega(S_\mathrm{n} - \dot{r}\tilde{S}_\lambda)\,\hat{\ell} \,,
\label{S-cross-v}
\\
\vec{\Sigma}\times\vec{v} 
&=& -v^2\tilde{\Sigma}_\ell\,\hat{n}
+ \dot{r}\tilde{\Sigma}_\ell\,\vec{v}
+ r\Omega(\Sigma_\mathrm{n} - \dot{r}\tilde{\Sigma}_\lambda)\,\hat{\ell} \,,
\label{Sigma-cross-v}
\end{eqnarray}
where $\Omega$ is the instantaneous orbital frequency, 
$\hat{\lambda} \equiv \hat{\ell}\times\hat{n}$,
and we have introduced the notation
\begin{eqnarray*}
&&
S_\ell = \vec{S}\cdot\hat{\ell} \,, \quad
S_\mathrm{n} = \vec{S}\cdot\hat{n} \,, \quad
S_\lambda = \vec{S}\cdot\hat{\lambda} \,,
\cr &&
\Sigma_\ell = \vec{\Sigma}\cdot\hat{\ell} \,, \quad
\Sigma_\mathrm{n} = \vec{\Sigma}\cdot\hat{n} \,, \quad
\Sigma_\lambda = \vec{\Sigma}\cdot\hat{\lambda} \,, 
\end{eqnarray*}
and
\begin{eqnarray*}
&&
\tilde{S}_\ell = S_\ell/(r\Omega) \,, \quad
\tilde{S}_\lambda = S_\lambda/(r\Omega) \,, \cr
&&
\tilde{\Sigma}_\ell =\Sigma_\ell/(r\Omega) \,, \quad
\tilde{\Sigma}_\lambda = \Sigma_\lambda/(r\Omega) \,.
\end{eqnarray*}
The PN expansion for $\Omega$ is only calculated in the case 
of quasi-circular orbits.
To avoid assuming quasi-circular orbits, 
we may replace $r\Omega$ in the above identities with the (exact) identity
\begin{eqnarray}
 r\Omega = |\hat{n}\times\vec{v}| = \sqrt{v^2 - \dot{r}^2} \,.
\end{eqnarray}
After using Eqs.~\eqref{n-cross-S}--\eqref{Sigma-cross-v} to eliminate 
the cross-products in Eq.~\eqref{direct_EOM} and recollecting terms,
one obtains an acceleration equation in the form of Eq.~\eqref{pEOM},
with
\begin{widetext}
\begin{subequations}
\begin{eqnarray} \label{EOM_ev}
\tilde{\mathcal{A}} &=& {\mathcal{A}} 
- \left(\frac{m}{r^2}\right)^{-1}
\left[
  \dot{r}\tilde{S}_\ell\,{\mathcal{C}}_{1} 
  + \dot{r}\tilde{\Sigma}_\ell\,{\mathcal{C}}_{2} 
  - v^2\tilde{S}_\ell\,{\mathcal{C}}_{3} 
  - v^2\tilde{\Sigma}_\ell\,{\mathcal{C}}_{4}
\right] \,,
\\
\tilde{\mathcal{B}} &=& {\mathcal{B}}
- \left(\frac{m}{r^2}\right)^{-1}
\left[
  - r\Omega\tilde{S}_\ell\,{\mathcal{C}}_{1} 
  - r\Omega\tilde{\Sigma}_\ell\,{\mathcal{C}}_{2} 
  + \dot{r}\tilde{S}_\ell\,{\mathcal{C}}_{3} 
  + \dot{r}\tilde{\Sigma}_\ell\,{\mathcal{C}}_{4}
\right] \,,
\\
\tilde{\mathcal{C}} &=& 
- \left(\frac{m}{r^2}\right)^{-1} r\Omega
\left[
  \tilde{S}_\lambda\,{\mathcal{C}}_{1} 
  + \tilde{\Sigma}_\lambda\,{\mathcal{C}}_{2} 
  + (S_\mathrm{n} - \dot{r}\tilde{S}_\lambda)\,{\mathcal{C}}_{3} 
  + (\Sigma_\mathrm{n} - \dot{r}\tilde{\Sigma}_\lambda)\,{\mathcal{C}}_{4}
\right] \,.
\end{eqnarray}
\end{subequations}
\end{widetext}
It is noted that this last equation is used in the case of quasi-circular orbits 
to compute the orbital plane precession frequency $\varpi$. 
Also we note that in the case when spins are 
aligned or anti-aligned, $\mathcal{C}$ vanishes identically (since
in that case $S_\mathrm{n}=\tilde{S}_\lambda=0$ 
and $\Sigma_\mathrm{n}=\tilde{\Sigma}_\lambda=0$).

\section{Spin Precession Equations}\label{sec:sEOM}

We overview the spin precession equations for two bodies
with spins that are unaligned with the orbital angular momentum.
As mentioned previously, these equations are needed
to complete the EOM.

There is no unique definition of center of mass in a
relativistic theory. The notion of spin inherits this ambiguity, which 
gives rise to non-physical degrees of freedom that need to be controlled.
This is done by imposing a ``spin supplementary condition'' (or SSC) that
eliminates the non-physical degrees of freedom. The SSC is imposed on 
the spin tensor, out of which are constructed a spin 4-vector and 
eventually a spin 3-vector, with just three physical degrees of freedom.
Following Refs.~\cite{Faye:2006gx, Blanchet:2006gy, Marsat:2012fn, 
Bohe:2012mr}, we use the Tulczyjew SSC~\cite{Kidder:1995zr}.
\begin{eqnarray}
S^{\mu\nu}p_{\nu} = 0 \,.
\end{eqnarray}

We define spin 3-vectors with conserved norm by $\vec{S}_\mathrm{a}$, where 
$\mathrm{a}=1,\,2$ is the label of the particles. In terms of these conserved 
norm spin vectors, the precession equations are written as
\begin{eqnarray} \label{B:388}
\frac{\mathrm{d}\vec{S}_\mathrm{a}}{\mathrm{d}t} = \vec{\Omega}_\mathrm{a} \times \vec{S}_\mathrm{a} \,,
\end{eqnarray}
where $\vec{\Omega}_\mathrm{a}$ with $\mathrm{a}=1,\,2$ are the precession 
vectors for mass 1 and 2, respectively. Each precession vector can be 
decomposed as
\begin{eqnarray}
\vec{\Omega}_\mathrm{a} 
= \vec{\Omega}_\mathrm{a,NS} 
+ \vec{\Omega}_\mathrm{a,SO}
+ \vec{\Omega}_\mathrm{a,SS} 
+ O{(SSS)} \,.
\label{OmegaSpin}
\end{eqnarray}
The precession vectors are only expanded to quadratic-in-spin order, 
$O{(SS)}$, because they get multiplied by a spin vector in the precession 
equation. Terms in the precession vectors at $O{(SSS)}$ would contribute
at $O{(SSSS)}$ in the precession equations, which is beyond the scope of 
this work. Each contribution to the precession vector is then decomposed
into a PN expansion of the form,
\begin{subequations}
\begin{eqnarray}
\vec{\Omega}_\mathrm{a,NS}
&=& \frac{1}{c^2}\vec{\Omega}_\mathrm{a,NS}^{(\mathrm{1PN})}
+ \frac{1}{c^4}\vec{\Omega}_\mathrm{a,NS}^{(\mathrm{2PN})} 
+ \frac{1}{c^6}\vec{\Omega}_\mathrm{a,NS}^{(\mathrm{3PN})} 
\cr &&
+ \frac{1}{c^7}\vec{\Omega}_\mathrm{a,NS}^{(\mathrm{3.5PN})} 
+ O\left(\frac{1}{c^8}\right) \,,
\label{OmegaSpin-NS}
\\
\vec{\Omega}_\mathrm{a,SO}
&=& \frac{1}{c^3}\vec{\Omega}_\mathrm{a,SO}^{(\mathrm{1.5PN})}
+ \frac{1}{c^5}\vec{\Omega}_\mathrm{a,SO}^{(\mathrm{2.5PN})} 
+ \frac{1}{c^7}\vec{\Omega}_\mathrm{a,SO}^{(\mathrm{3.5PN})} 
\cr &&
+ O\left(\frac{1}{c^8}\right) \,,
\label{OmegaSpin-SO}
\\
\vec{\Omega}_\mathrm{a,SS}
&=& \frac{1}{c^6}\vec{\Omega}_\mathrm{a,SS}^{(\mathrm{3PN})}
+ O\left(\frac{1}{c^8}\right) \,.
\label{OmegaSpin-SS}
\end{eqnarray}
\end{subequations}
Explicit expressions for $\vec{\Omega}_1$ and $\vec{\Omega}_2$ 
are given in Ref.~\cite{Bohe:2012mr}.

\section{The Taylor T4 Method}\label{sec:T4_method}

Taylor T4 is an \textit{adiabatic} approximation~\footnote{An adiabatic
approximation means that we do not consider the change in any quantity 
that is smaller than one orbit, i.e., the inspiral of the BHs will not 
affect the orbit. As such, since we consider spin 
precession on a different timescale than the orbital timescale, these 
approximations do not have spin precession built into them,
and are only for aligned spins.}
that evolves the phase and frequency of the BBH. There are two parts 
to this method. The first is the phase evolution of the binary,
which we will detail here,
and the second is the calculation of the gravitational radiation from the orbital 
phasing of the binary. 

All of the formulas of energy and flux can be written in terms of the frequency 
variable $v\equiv (M \mathrm{d}\phi/\mathrm{d}t)^{1/3}$, and we 
start with the basic definition of energy conservation, namely:
\bea
\frac{\mathrm{d}E(v)}{\mathrm{d}t} + \mathcal{F}(v) + \dot{M}(v) = 0 \,,
\eea
where the time rate of change of the energy is equal to the flux leaving the 
system plus the mass rate of change due to GW absorption.
In practice, since this BH absorption is a relatively tiny
contribution~\cite{Isoyama:2017tbp}
(although 2.5PN order relative to the leading flux),
we neglect it for the rest of this analysis, bringing our conservation statement to
\bea \label{Ebalance}
\frac{\mathrm{d}E(v)}{\mathrm{d}t} = -\mathcal{F}(v) \,. 
\eea

We start with the energy balance equation~\eqref{Ebalance},
integrate it to find $v(t)$, and therefore $\Omega(t)$, since 
$\mathrm{d}\phi/\mathrm{d}t = \Omega$. Taylor T4 explicitly takes the rational fraction,
\bea \label{T4_ev}
\frac{\mathrm{d}v}{\mathrm{d}t} = - \frac{\mathcal{F}}{\mathrm{d}E/\mathrm{d}v} \,, 
\eea
along with 
\bea
\frac{\mathrm{d}\phi}{\mathrm{d}t} =  \frac{v^3}{M} \,,
\eea
re-expands the fraction as a 
consistent PN series, then numerically integrates to obtain $v(t)$ and $\phi(t)$. 

Full expressions of $E(v)$, ${\cal F}(v)$, and $\mathrm{d}E/\mathrm{d}v$ that we use are given
in Ref.~\cite{Brown:2007jx}.
These equations assume a quasi-circular orbit for the trajectory. 
Taylor T4 approximants integrate the energy balance equations and have 
been thoroughly explored and developed. Further approximations can be 
constructed in the frequency domain (the Taylor F-approximants). 
These methods have been compared,
see Ref.~\cite{Buonanno:2009PhRvD..80h4043B}.

\clearpage

\begin{widetext}
\section{Tabulated results on overlap calculations}

\begin{center}
\begin{table*}[h!] 
\resizebox{\textwidth}{!}{
\begin{tabular}{|c|c|c|c|c|c|c|c|c|c|}
\hline
\multicolumn{2}{|c|}{Separation} &  & \multicolumn{2}{c|}{}
& \multicolumn{3}{c|}{Duration} 
& \multicolumn{2}{c|}{Overlaps $O_{{\rm T4}-{\rm EOM}}$} \\ \hline
$r_{\rm init}$ $[M]$ & $r_{\rm fin}$ $[M]$ & $q$ & $\chi_1$ & $\chi_2$ & $t$ [M] & orbits & $\mathrm{d} t$ [s] & ${\rm T4_{consistent}}$ & ${\rm T4_{high}}$ \\ \hline \hline
100 & 99 & 1 & 0 & 0 & 318429.8 & $\sim 50$ & $1.09061\times 10^{-3}$ & 0.99999065 & 0.999990644 \\ \hline
100 & 99 & 1 & 0.3 & 0.3 & 318429.8 & $\sim 50$ & $1.04527\times 10^{-3}$ & 0.99993352 & 0.999933512 \\ \hline
100 & 99 & 1 & 0.6 & 0.6 & 318429.8 & $\sim 50$ & $1.04527\times 10^{-3}$ & 0.99983625 & 0.999836244 \\ \hline
100 & 99 & 1 & 0.9 & 0.9 & 318429.8 & $\sim 50$ & $1.04527\times 10^{-3}$ & 0.99971069 & 0.999710688 \\ \hline
100 & 99 & 1 & 0.3 & -0.3 & 318429.8 & $\sim 50$ & $1.04527\times 10^{-3}$ & 0.99998949 & 0.999989489 \\ \hline
100 & 99 & 1 & 0.6 & -0.6 & 318429.8 & $\sim 50$ & $1.04527\times 10^{-3}$ & 0.9999856287 & 0.9999856265 \\ \hline
100 & 99 & 1 & 0.9 & -0.9 & 318429.8 & $\sim 50$ & $1.04527\times 10^{-3}$ & 0.99997797 & 0.9999778678 \\ \hline
100 & 99 & 1 & 0.9 & -0.5 & 318429.8 & $\sim 50$ & $1.04527\times 10^{-3}$ & 0.9999415165 & 0.9999415133 \\ \hline
100 & 99.1 & 2 & 0 & 0 & 318480.2 & $\sim 50$ & $1.045439\times 10^{-3}$ & 0.99999283 & 0.999992828 \\ \hline
100 & 99.7 & 10 & 0 & 0 & 318701.62 & $\sim 50$ & $1.04617\times 10^{-3}$ & 0.9999991669 & 0.9999991668 \\ \hline
100 & 99.95 & 100 & 0 & 0 & 318833.8 & $\sim 50$ & $1.0466\times 10^{-3}$ & 0.9999999893 & 0.9999999893 \\ \hline \hline
50 & 47.2 & 1 & 0 & 0 & 114043.4 & $\sim 50$ & $3.90593\times 10^{-4}$ & 0.994904572 & 0.994898471 \\ \hline
50 & 47.2 & 1 & 0 & 0 & 114049.1 & $\sim 50$ & $2.246257\times 10^{-4}$ & 0.994797482 & 0.994791317 \\ \hline
50 & 47.3 & 1 & 0.3 & 0.3 & 114043.4 & $\sim 50$ & $3.74357\times 10^{-4}$ & 0.983693663 & 0.983675238 \\ \hline
50 & 47.3 & 1 & 0.6 & 0.6 & 114043.4 & $\sim 50$ & $3.74357\times 10^{-4}$ & 0.969583498 & 0.969553474 \\ \hline
50 & 47.4 & 1 & 0.7 & 0.7 & 114043.4 & $\sim 50$ & $3.74357\times 10^{-4}$ & 0.981885223 & 0.981867878 \\ \hline
50 & 47.4 & 1 & 0.8 & 0.8 & 114043.4 & $\sim 50$ & $3.74357\times 10^{-4}$ & 0.991724021 & 0.991717488 \\ \hline
50 & 47.4 & 1 & 0.85 & 0.85 & 114043.4 & $\sim 50$ & $3.74357\times 10^{-4}$ & 0.992601112 & 0.992597499 \\ \hline
50 & 47.4 & 1 & 0.9 & 0.9 & 114043.4 & $\sim 50$ & $3.74357\times 10^{-4}$ & 0.990197785 & 0.990190154 \\ \hline
50 & 47.4 & 1 & 0.95 & 0.95 & 114043.4 & $\sim 50$ & $3.74357\times 10^{-4}$ & 0.984804119 & 0.984791162 \\ \hline
50 & 47.2 & 1 & 0.3 & -0.3 & 114043.4 & $\sim 50$ & $3.74357\times 10^{-4}$ & 0.995280388 & 0.995274373 \\ \hline
50 & 47.2 & 1 & 0.6 & -0.6 & 114043.4 & $\sim 50$ & $3.74357\times 10^{-4}$ & 0.9971254502 & 0.997121275 \\ \hline
50 & 47.2 & 1 & 0.9 & -0.9 & 114043.4 & $\sim 50$ & $3.74357\times 10^{-4}$ & 0.9988241423 & 0.998822279 \\ \hline
50 & 47.3 & 1 & 0.9 & -0.5 & 114043.4 & $\sim 50$ & $3.74357\times 10^{-4}$ & 0.985182006 & 0.985164755 \\ \hline
50 & 47.5 & 2 & 0 & 0 & 114095.5 & $\sim 50$ & $3.74528\times 10^{-4}$ & 0.99467073 & 0.994667208 \\ \hline
50 & 49.1 & 10 & 0 & 0 & 114299.9 & $\sim 50$ & $3.75199\times 10^{-4}$ & 0.999489566 & 0.999489472 \\ \hline
50 & 49.9 & 100 & 0 & 0 & 114395.1 & $\sim 50$ & $3.75512\times 10^{-4}$ & 0.999994205 & 0.999994204 \\ \hline \hline
20 & 16 & 1 & 0 & 0 & 11965 & $\sim 20$ & $3.92762\times 10^{-5}$ & 0.782932526 & 0.775227367 \\ \hline
20 & 18.2 & 1 & 0 & 0 & 5982.5 & $\sim 10$ & $1.963812\times 10^{-5}$ & 0.917453743 & 0.916892613 \\ \hline
20 & 16.5 & 2 & 0 & 0 & 11979.2 & $\sim 20$ & $3.93229\times 10^{-5}$ & 0.827044247 & 0.821830277 \\ \hline
20 & 18.8 & 10 & 0 & 0 & 12050.8 & $\sim 20$ & $3.95579\times 10^{-5}$ & 0.972718171 & 0.97255266 \\ \hline
20 & 19.9 & 100 & 0 & 0 & 12087.1 & $\sim 20$ & $3.96769\times 10^{-5}$ & 0.99919055 & 0.999190336 \\ \hline
\end{tabular}}
\caption{The overlaps between Taylor T4 and EOM methods, documented
for different separations, mass ratios ($q$), and
aligned dimensionless spin values ($\chi_1$ and $\chi_2$), 
at both the consistent T4 flux order and the highest T4 fluxes available. 
There are several interesting lines to pay attention to: as the spins increase
in alignment, there is an inflection point around $\chi_1 = \chi_2 = 0.9$,
where the overlap drops then increases again. In addition, at $50M$ separation,
we increase the resolution of the timestep to probe the effects
of numerical resolution on the overlap calculations.
We find no effect on the overlap of the waveforms due to the timestep resolution.}
\label{tab:T4_EOM_overlap_comparisons}
\end{table*}
\end{center}
\end{widetext}

\bibliographystyle{apsrev}
\bibliography{./references}

\end{document}